\documentclass[twocolumn, prx, superscriptaddress,notitlepage]{revtex4-2}
\usepackage{graphicx}
\usepackage{dcolumn}
\usepackage{bm}
\usepackage[usenames,dvipsnames]{color}
\usepackage[most]{tcolorbox}
\usepackage{multirow}
\usepackage{gensymb}
\usepackage[normalem]{ulem}
\usepackage{CJK}
\usepackage{comment}
\usepackage[colorlinks, linkcolor=blue,anchorcolor=blue,citecolor=blue,urlcolor=blue]{hyperref}
\usepackage{amssymb}
\usepackage{pifont}
\usepackage{physics}
\usepackage{natbib}
\usepackage{dsfont}
\usepackage{mathrsfs}
\newcommand{\ave}[1]{\left\langle #1\right\rangle}

\newcommand{\sz}{\sigma_z}

\newcommand{\Black}[1]{\textcolor[rgb]{0,0,0}{#1}}

\newcommand{\lv}[1]{\textcolor[rgb]{0,0,0}{#1}}

\usepackage{color,soul}

\begin{document}
\begin{CJK*}{UTF8}{}
\title{Digital noise spectroscopy with a quantum sensor}

\author{Guoqing Wang \CJKfamily{gbsn}(王国庆)}\email[]{gq\_wang@mit.edu}
\affiliation{
   Research Laboratory of Electronics, Massachusetts Institute of Technology, Cambridge, MA 02139, USA}
\affiliation{
   Department of Nuclear Science and Engineering, Massachusetts Institute of Technology, Cambridge, MA 02139, USA}

\author{Yuan Zhu}
\affiliation{
   Research Laboratory of Electronics, Massachusetts Institute of Technology, Cambridge, MA 02139, USA}
\affiliation{
   Department of Nuclear Science and Engineering, Massachusetts Institute of Technology, Cambridge, MA 02139, USA}
\affiliation{
   Department of Electrical Engineering and Computer Science, Massachusetts Institute of Technology, Cambridge, MA 02139, USA}

\author{Boning Li}
\affiliation{
   Research Laboratory of Electronics, Massachusetts Institute of Technology, Cambridge, MA 02139, USA}
\affiliation{Department of Physics, Massachusetts Institute of Technology, Cambridge, MA 02139, USA}
   
\author{\mbox{Changhao Li}}
\thanks{Current address: Global Technology Applied Research, JPMorgan Chase, New York, NY 10017 USA}
\affiliation{
   Research Laboratory of Electronics, Massachusetts Institute of Technology, Cambridge, MA 02139, USA}
\affiliation{
   Department of Nuclear Science and Engineering, Massachusetts Institute of Technology, Cambridge, MA 02139, USA}

\author{Lorenza Viola}
\affiliation{\mbox{Department of Physics and Astronomy, Dartmouth College, 6127 Wilder Laboratory, Hanover, NH 03755, USA}}

\author{Alexandre Cooper}
\affiliation{Institute for Quantum Computing, University of Waterloo, Waterloo, ON N2L 3G1, Canada}

\author{Paola Cappellaro}\email[]{pcappell@mit.edu}
\affiliation{
   Research Laboratory of Electronics, Massachusetts Institute of Technology, Cambridge, MA 02139, USA}
\affiliation{
   Department of Nuclear Science and Engineering, Massachusetts Institute of Technology, Cambridge, MA 02139, USA}
\affiliation{Department of Physics, Massachusetts Institute of Technology, Cambridge, MA 02139, USA}

\begin{abstract}
We introduce and experimentally demonstrate a quantum sensing protocol to sample and reconstruct the auto-correlation of a noise process 
using a
single-qubit sensor \lv{under digital control modulation}. This Walsh noise spectroscopy method exploits simple sequences of spin-flip pulses to generate a complete basis of digital filters that directly sample the power spectrum of the \lv{target noise}
in the sequency domain -- from which the auto-correlation function 
in the time domain, as well as the power spectrum in the frequency domain, can be reconstructed using 
linear transformations. Our method, \lv{which can also be seen as an implementation of frame-based noise spectroscopy}, solves the fundamental difficulty in sampling continuous functions with digital filters by introducing a transformation that relates the arithmetic and logical time domains. In comparison to \lv{standard, frequency-based dynamical-decoupling}
noise spectroscopy protocols, the accuracy of our method is only \Black{limited by the sampling and discretization in the time space} and can be easily improved, even under limited evolution time due to decoherence and hardware limitations. Finally, we experimentally reconstruct the auto-correlation function of the effective magnetic field produced by the nuclear-spin bath on the electronic spin of a single nitrogen-vacancy center in diamond, discuss practical limitations of the method, and avenues for \lv{further} improving the reconstruction accuracy.
\end{abstract}

\maketitle

\end{CJK*}	

\section{Introduction}

Characterizing fluctuating fields and environmental noise in quantum systems is an essential task not only for fundamental physics, \lv{for instance in the context of elucidating} the interaction mechanisms in condensed matter systems~\cite{yang_quantum_2017,dobrovitski_decay_2009}, but also for quantum applications such as building high-fidelity quantum devices robust against environmental noise~\cite{de_lange_universal_2010,biercuk_optimized_2009,alvarez_measuring_2011,suter_colloquium_2016}. 
In recent years, a variety of noise reconstruction techniques \lv{-- broadly referred to as quantum noise spectroscopy (QNS) --have been proposed and experimentally} demonstrated in different platforms~\cite{szankowski_environmental_2017}, including trapped ions~\cite{kotler_single-ion_2011, Kim2017, frey2017, frey2020}, superconducting qubits~\cite{bylander_noise_2011,sung2019,von_lupke_two-qubit_2020}, quantum dots~\cite{medford_scaling_2012,dial_charge_2013,chan2018}, and spin defects in diamond~\cite{staudacher_nuclear_2013,bar-gill_suppression_2012,romach2015}.

The 
protocols \lv{which have been most commonly employed to date},
inspired by classical signal processing,  perform spectral reconstruction in the frequency domain by measuring the decoherence of a qubit sensor subjected to sequences of spin-flip pulses~\cite{szankowski_environmental_2017,yuge_measurement_2011,alvarez_measuring_2011}, also known as a dynamical decoupling sequences because of their noise filtering capability. Whereas these protocols claim a simple intuitive formalism, the exact digital filters that they produce are a product of trigonometric functions -- effectively converting the reconstruction problem into a non-linear inversion problem that necessitates numerical deconvolution algorithms~\cite{bar-gill_suppression_2012}. Although approximating \lv{such frequency-domain digital filter function (FF)} with a delta-like function centered at some specific frequency can solve this problem in principle, such an approximation \lv{is not well controlled in general and fails to address the fundamental
challenge of sampling functions of a continuous (frequency or time) variable with digital filters. In addition, it is practically difficult} to mitigate the reconstruction error introduced by the delta-filter approximation when the sampling time is limited due to short coherence times and hardware limitations.

In this work, we propose and experimentally demonstrate the benefits of using digital (Walsh) QNS to sample and reconstruct stochastic fields using quantum sensors. The Walsh QNS method uses a sequence of coherent control pulses acting on a single-qubit sensor to generate a complete set of digital filters. Each filter encodes a specific coefficient of the \emph{logical} power spectrum in the decoherence rate of the sensor. By measuring the decoherence exponent using a complete set of Walsh modulation sequences, an estimate of the logical power spectrum  can be reconstructed, from which the auto-correlation function and power spectrum of the noise can be directly computed using a sequence of linear transformations. \lv{Notably, our method shares important points of contact with a general {\em frame-based} approach \cite{Chalermpusitarak21}, which was recently shown to afford a resource-efficient characterization in the presence of finite control resources. Here, by}
analytically and numerically calculating the accuracy in reconstructing the noise spectrum, we show that Walsh QNS provides a comparable or even better alternative to standard dynamical decoupling-based methods \lv{operating in the frequency domain}, especially when reconstructing the auto-correlation function of the noise. We then experimentally demonstrate our method by characterizing the environmental noise of a single nitrogen-vacancy (NV) center in diamond, resulting from interactions with its $^{13}$C nuclear spin bath. We finally discuss practical limitations of the method and outline strategies to further improve its performance. 

\section{Theory}
\subsection{Walsh spectroscopy method}

Different quantum systems \lv{may be exposed to} quite different noise sources, such as charge and flux fluctuations in superconducting circuits and magnetic spin baths in solid-state systems~\cite{szankowski_environmental_2017}. Although the microscopic decoherence mechanism  may be due to an environment that is quantum \lv{(non-commuting)} in nature~\cite{yang_quantum_2017}, the noise process can often be approximated by a classical Gaussian process~\cite{szankowski_environmental_2017}. Here, we focus on reconstructing a stationary Gaussian noise that leads to pure dephasing of a single-qubit sensor. \lv{In a frame that rotates with the qubit frequency}, 
we can assume that the qubit is coupled to a longitudinal noise field $\omega(t)$ by the Hamiltonian 
\begin{equation}
    \mathcal{H}=\frac{\omega(t)}{2}\sz,
\end{equation}
where, as usual, $\sz$ denotes the $z$ Pauli matrix.
The noise-induced dephasing can be observed by performing Ramsey interferometry while modulating the state of the qubit with a sequence of $m$ instantaneous spin-flip ($\pi$) pulses. 
The dynamical phase acquired by the sensor is
\begin{align}\label{eq:phase}
\varphi_m(T)&=\int_0^T \omega(t)f_m(t)dt,
\end{align}
where $f_m(t)=\pm1$ is the modulation induced by the $\pi$ pulses. Assuming a stationary and zero-mean Gaussian distribution of the stochastic noise field $\omega(t)$, the decoherence exponent $\chi$ of the measured signal is given by
\begin{align}
\chi_m(T)\!&=\!\frac{\langle\varphi_m^2(T)\rangle}2\!\nonumber \\&=\!\frac12\int_0^T\!\!\!\int_0^T\!\!\!\!  \langle\omega(t_1)\omega(t_2)\rangle f_m(t_1)f_m(t_2)dt_1 dt_2.
\label{eq:chi_theory}
\end{align}
The qubit decay is thus determined by the noise auto-correlation function, $G(t_1,t_2)\equiv\langle\omega(t_1)\omega(t_2)\rangle$, which simplifies to \lv{an even function of the time lag, 
$G(t_1,t_2) \equiv G(t_1-t_2)= G(t_2-t_1)$ under the assumption of stationarity. Accordingly,} reconstructing the auto-correlation suffices to fully characterize the noise process. 

\begin{figure*}[htb]
\centering \includegraphics[width=0.95\textwidth]{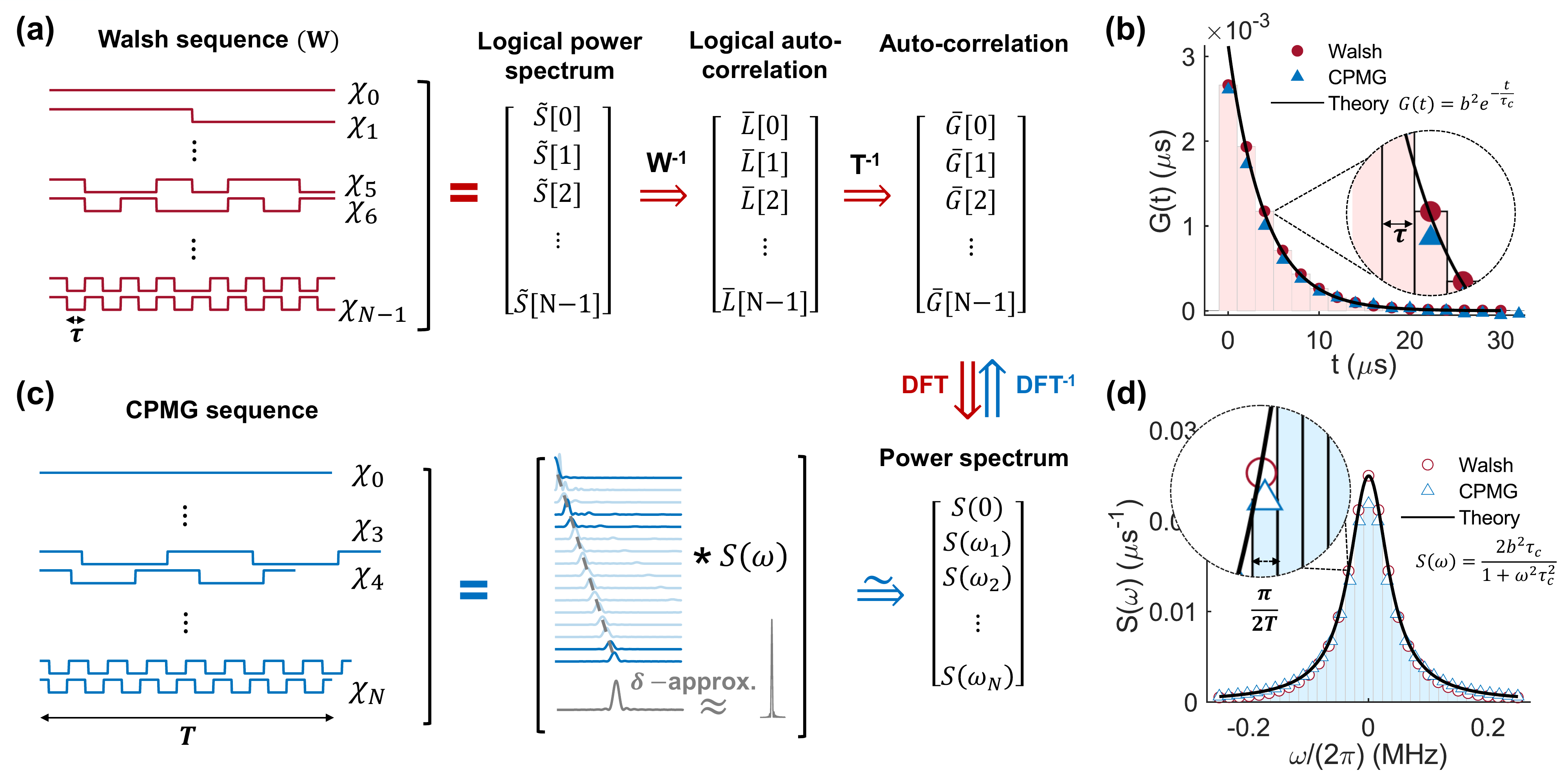}
\caption{\label{Fig_WalshCPMG_main}\textbf{Walsh noise spectroscopy protocol}.
(a)~A complete set of $N$ Walsh sequences $\{w_m(t)\}$ samples the coefficients of the logical power spectrum $\tilde{S}(m)$ in the sequency domain up to order $N=2^n$
by measuring the decoherence exponents $\chi_m(T)$ after a sampling time $T$. The logical power spectrum $\tilde{S}[m]$ is converted into the logical auto-correlation function $\overline L[j]$ by performing the inverse Walsh transform; $\overline L[j]$ is then converted into the auto-correlation function $\overline G(t)$ by performing the (pseudo-)inverse shuffling transformation $T^{-1}$, which is finally converted into the power spectrum $S(\omega)$ by computing the discrete Fourier transform (DFT).
(b)~The Walsh sampling and reconstruction procedure results in an $N$-point approximation of the auto-correlation function of $G(t)$ over $[0,T)$ in the time domain, which is converted in the power spectrum $S(\omega)$ in the frequency domain.
(c)~\lv{Standard comb-based protocol, where a} set of $N+1$ CPMG sequences directly samples the power spectrum $S(\omega_k)$ in the frequency domain by generating frequency FFs whose peak-frequency increases linearly with the sequence indexing number. The auto-correlation function is recovered by computing the inverse DFT~(see Supplemental Materials~\cite{SOM} for details on the sequences). (d)~The CPMG sampling and reconstruction procedure results in an approximation of the power spectrum of $S(\omega)$ in the frequency domain, which can be converted to the auto-correlation in the time domain.
}
\end{figure*}

\lv{In order to achieve digital noise characterization, we choose
modulating functions $f_m$ given by the complete set of Walsh functions $\{w_m\}_{m=0}^{N-1}$ of order $N=2^n$~\cite{robinson_logical_1972}. In addition to also providing a paradigmatic instance of a digital frame \cite{frames_intro, Chalermpusitarak21},} the Walsh basis is convenient for its ease of practical implementation.
The ``sequency'' $m$ denotes the number of spin-flip pulses of the corresponding Walsh sequence, in analogy to the ``frequency'' of trigonometric functions in Fourier analysis of analog signals. As shown in Fig.~\ref{Fig_WalshCPMG_main}(a), the Walsh modulation functions $w_m$ are piecewise-continuous functions defined over $N$ contiguous intervals of length $\tau= T/N$. Equivalently, the binary values of the $m^{th}$ Walsh modulation function are given by the $m^{th}$ row of the $N\times N$ Walsh matrix $W$, such that $w_m(t/\tau)=W[m,j]$, where $t\in[j\tau,(j+1)\tau)$. Walsh functions are intrinsically compatible with the available quantum control \lv{hardware, supporting} microwave or optical pulses and time discretization~\cite{robinson_logical_1972}, and have been used to design optimal dynamical decoupling for coherence protection \lv{in the presence of both dephasing~\cite{hayes_reducing_2011} and general multi-axis noise~\cite{qi_optimal_2017-1}}. 

When combined with a qubit sensor, we  previously showed that Walsh functions can be used to reconstruct the temporal profile of deterministic time-varying magnetic fields~\cite{cooper_time-resolved_2014, magesan_2013}, by measuring a finite set of coherent phases $\{\varphi_m\}$~(see Eq.~\eqref{eq:phase}). The problem of sampling a stochastic field to reconstruct its auto-correlation $G$ from a set of decay exponents $\{\chi_m\}$ \lv{is a more challenging task}. 
The core difficulty stems from the exponent $\chi_m$ being given by the overlap between two functions naturally defined in {\em two} different domains: the auto-correlation function in the \emph{arithmetic} domain, and the Walsh functions in the \emph{logical} (\emph{dyadic}) domain~\cite{robinson_logical_1972},
\begin{equation}
\chi_m(T)=\frac12\iint G(t_1-t_2)w_m(\lfloor \frac{t_1}{\tau}\rfloor \oplus \lfloor \frac{t_2}{\tau}\rfloor)dt_1dt_2,
\label{eq:Rw}
\end{equation}  
where the floor symbol ``$\lfloor\cdot\rfloor$" means taking the closest smaller integer. Here we mathematically used the composition property of Walsh functions under multiplication, $w_m(h)w_m(j)=w_m(h\oplus j)$, with $(h\oplus j)$ denoting the binary addition modulo $N$. \lv{While a general solution to the problem of representing dynamical overlap integrals, such as the one determining $\chi_m(T)$, in a way that is directly tied to the available control resources is provided by the frame-based formalism developed in \cite{Chalermpusitarak21},} in the following we show that this difficulty can be elegantly overcome by using a sequence of simple, linear transformations. \lv{We further discuss the relationship to the frame approach in Appendix~\ref{app:frames}.}

First, to connect the arithmetic domain to the logical domain, we apply the change of variables, $u(t_1,t_2)=(t_1-t_2)\in(-T,T)$, $v(t_1,t_2)=\lfloor t_1/\tau\rfloor \oplus \lfloor t_2/\tau\rfloor\in[0,N)$, with an associated {\em shuffling transformation} $T(u,v)$, in analogy to a Jacobian matrix. Then, the decay exponent in Eq.~\eqref{eq:Rw} becomes
\begin{eqnarray}
\chi_m(T)&=&\frac12\iint G(u)\,T(u,v)\,w_m(v) \,du\,dv
\label{eq:RTw_main}.
\end{eqnarray} 
Carrying \lv{out} the integration over the arithmetic domain ($u$), with $L(v)\equiv \int G(u)T(u,v)du$ yields our key result, 
\begin{eqnarray}
\chi_m(T)~&=&\frac12\int L(v)w_m(v)dv\equiv\tilde{S}_T(m)\label{eq:Sm}.
\end{eqnarray}
\lv{This} shows that the decoherence exponent \lv{$\chi_m(T)$ under Walsh modulation with sequency $m$} is equivalent to the finite-time logical power spectrum $\tilde{S}_T(m)$ in the logical domain. In analogy to the arithmetic power spectrum $S(\omega)$ being given by the continuous Fourier transform of the auto-correlation function $G(t)$, here the logical power spectrum  $\tilde{S}_T(m)$ is given by the Walsh transform 
of the logical auto-correlation function $L(v)$, defined as 
\begin{align}
     L(v)\equiv \tau \int \langle \omega(\lfloor \frac{t}{\tau}\rfloor )\omega(\lfloor \frac{t}{\tau}\rfloor\oplus v) \rangle dt, \quad v\in[0,N).
   \label{eq:LN}
\end{align}
The auto-correlation function can thus be obtained from the logical power spectrum by applying the inverse Walsh transform and the shuffling transformation. 

To simplify the explicit calculations under finite values of $\tau$ and $N$, we introduce the discretized auto-correlation function in the arithmetic domain,
\begin{equation}
    \overline{G}_N[j-k]\equiv \frac{N^2}{T^2}\int_{\frac{kT}{N}}^{\frac{(k+1)T}{N}}\!\!\int_{\frac{jT}{N}}^{\frac{(j+1)T}{N}}\!\!\!\!G(t_1-t_2)dt_1dt_2, 
    \label{eq:Gapprox}
\end{equation}
as well as the dyadic time average of the logical auto-correlation, 
\begin{equation}
\overline L_N[j] \equiv \frac{N}{T^2}\int_{j}^{j+1}\!L(v)dv.    
\end{equation}
\lv{With these definitions, we may} rewrite 
Eq.~(\ref{eq:Sm}) and the relation between $L$ and $G$ as discrete sums:
\begin{eqnarray}
\widetilde S_T[m] &=&\frac{T^2}{2N}\sum_{j=0}^{N-1}W[m,j]\overline L_N[j],\label{eq:Smsum}\\
\overline L_N[j] &=&\frac1N \sum_{k=0}^{N-1}  \overline G_N[j\oplus k -k]\\
~&=&\frac1N \sum_{k=0}^{N-1} T_N[j,k]\overline G_N[k],\label{eq:LNsum}
\end{eqnarray}
where $T_N[j,k]$ is the matrix representation of the shuffling transformation,  which can be constructed recursively~\cite{SOM, robinson_logical_1972} or using convolution products of Walsh functions~(see Appendix~\ref{App_Theory} for a detailed derivation).

The logical auto-correlation function $\overline L_N$ can thus be simply obtained by sampling the coefficients of the logical power spectrum $\tilde{S}_T$.
\lv{In turn, this entails} measuring the decoherence exponents $\{\chi_m\}$ with a set of $N$ Walsh sequences $\{w_m(t)\}_{m=0}^{N-1}$ and computing the inverse Walsh transform (here the Walsh matrix $W^{-1}$), 
\begin{equation}
\overline L_N[j]=\frac{2N}{T^2}\sum_{m=0}^{N-1}W^{-1}[j,m]\tilde S_T[m].
\end{equation}
The average auto-correlation is finally obtained by applying the inverse of the linear shuffling transformation in Eq.~(\ref{eq:LNsum}). \lv{This yields}
\begin{equation}
 \overline G_N[j]\!=\!\!\sum_{k=0}^{N-1}\! \frac2{2^{H(k)}2^{\delta(k,0)}}T_N^{-1}[k,j]\overline L_N[k],
\end{equation}
where $H(k)$ is the Hamming weight of $k$ and $T_N^{-1}$ is the (pseudo-)inverse of the shuffling matrix.

\lv{We stress that, while the logical power spectrum $\tilde{S}_T(m)$ can be equated with the decoherence exponent, its relationship with the usual 
frequency-domain power spectrum $S(\omega)$ is not trivial}. The latter can be recovered by discretizing the continuous-time auto-correlation $G(t_j)$ into its time-averaged representation $\overline G[j]$ at discrete sampling times (see Eq.~\eqref{eq:Gapprox}), and applying the Fourier transform to obtain estimates $S(\omega_k)$, with $\omega_k=\frac{k\pi N}{T(N-1)}$, $k=0,\cdots,N-1$~\footnote{More precisely, 
\lv{thanks to the fact that the auto-correlation function is even, $G(t)=G(-t)$,} 
we obtain the discretized noise spectrum $S(\omega_k)$ by applying the discrete Fourier transform (DFT)~\cite{SOM}.}.

\subsection{Comparison to comb-based noise spectroscopy methods}

The most common QNS method, \lv{based on engineering a ``comb'' in frequency space through periodic sequence repetition}~\cite{alvarez_measuring_2011, szankowski_environmental_2017}, attempts to directly reconstruct the noise spectrum $S(\omega)$ at desired frequencies (Fig.~\ref{Fig_WalshCPMG_main}b).
Indeed, \lv{thanks to stationarity}, we can rewrite Eq.~\eqref{eq:chi_theory} by replacing the noise auto-correlation with its Fourier transform, which corresponds to the noise spectrum (see Appendix~\ref{App_Theory}). The resulting decoherence exponent  
can then be described as the overlap in the frequency domain between the noise spectrum and a FF \cite{szankowski_environmental_2017,Biercuk_2011,pazsilva2014}:
\begin{equation}
    \chi_m(T)=\frac12\int_{-\infty}^{+\infty}\frac{d\omega}{2\pi}S(\omega) \lv{|F_m(\omega, T)|^2},
\end{equation}
where the FF $F_m(\omega,T)$ is associated with the modulation function $f_m(t)$ through a \lv{finite-time} Fourier transform, $|F_m(\omega,T)|^2=|\int_0^Tf_m(t)e^{i\omega t}dt|^2$.

When \lv{a base sequence of spin-flip $\pi$ pulses of duration $\tau_b$, e.g., a two-pulse CPMG sequence with an inter-pulse delay $\tau$~\cite{carr_effects_1954,meiboom_modified_1958},
is periodically repeated in time,} the FF can be approximated by a weighted sum over a series of Dirac delta functions~\cite{alvarez_measuring_2011,yuge_measurement_2011,szankowski_accuracy_2018}, namely, 
\begin{equation}
    |F(\omega)|^2\approx\sum_{k=0}^{\infty}\frac{8T}{\pi^2}\frac{1}{(2k+1)^2}\,\delta\left(\omega-\frac{(2k+1)\pi}{\tau}\right).
\end{equation}
Then the decoherence exponent is dominated by the noise components at the \lv{base frequency $\omega_0\equiv 2\pi/\tau_b =\pi/\tau$} and its odd harmonics:
\begin{equation}
    \chi (T)=T\frac{4}{\pi^2}\sum_{k=0}^{+\infty}\frac{1}{(2k+1)^2}\,S\left(\frac{(2k+1)\pi}{\tau}\right),
    \label{eq:CPMG_chi}
\end{equation}
from which the noise spectrum $S(\omega)$ can be reconstructed by tuning the value of $\tau$ \lv{and truncating the above series to a maximum finite number of harmonics, $k_{\text{max}}$}. As experimentally demonstrated in different platforms~\cite{yuge_measurement_2011,alvarez_measuring_2011,bylander_noise_2011}, this method is powerful for solving certain operational problems, especially sampling the spectrum at selected  or  high-frequencies, However, the ``delta-filter'' approximated method lacks in generality, as hinted by the fact that periodic sequences are a subset of the Walsh basis \lv{(in particular, $m=2^n$ for CPMG and $m=2^n-1$ for PDD \cite{hayes_reducing_2011}).} 
In addition, for the approximation to be valid, the noise spectrum around the \lv{``resonance frequencies'' $(2k+1)\omega_0$ should have small variations with respect to the filter peak width}, which is set by the total sampling time, $T$. In other words, the sampling time $T$ has to be much larger than the noise correlation time $\tau_c$. 
Suppressing this error requires either solving the non-trivial inversion problem using the exact FF or increasing the measurement time $T$, often beyond the system coherence time, leading to low-accuracy estimates of $\chi_m(T)$~\footnote{\lv{See also Sec.\,V.B of Ref.\,\cite{norris_qubit_2016} for an expanded discussion,
applicable to more general frequency comb-based
protocols employing a set of different base sequences.}}. More practically, a long time $T$ also degrades the sensitivity and efficiency of noise reconstruction. Besides, the experimental apparatus might limit the length of achievable pulse sequences. 

In contrast, Walsh-based \lv{QNS -- and frame-based QNS more generally \cite{Chalermpusitarak21} --} enable a {\em direct} reconstruction of the noise auto-correlation, and the dominant sources of errors (sampling in the time domain and average auto-correlation approximation) are quite different. It is thus critical to quantitatively assess the performance of both methods and compare strengths and weaknesses for different noise properties. 

\begin{figure*}[htbp]
\centering \includegraphics[width=0.82\textwidth]{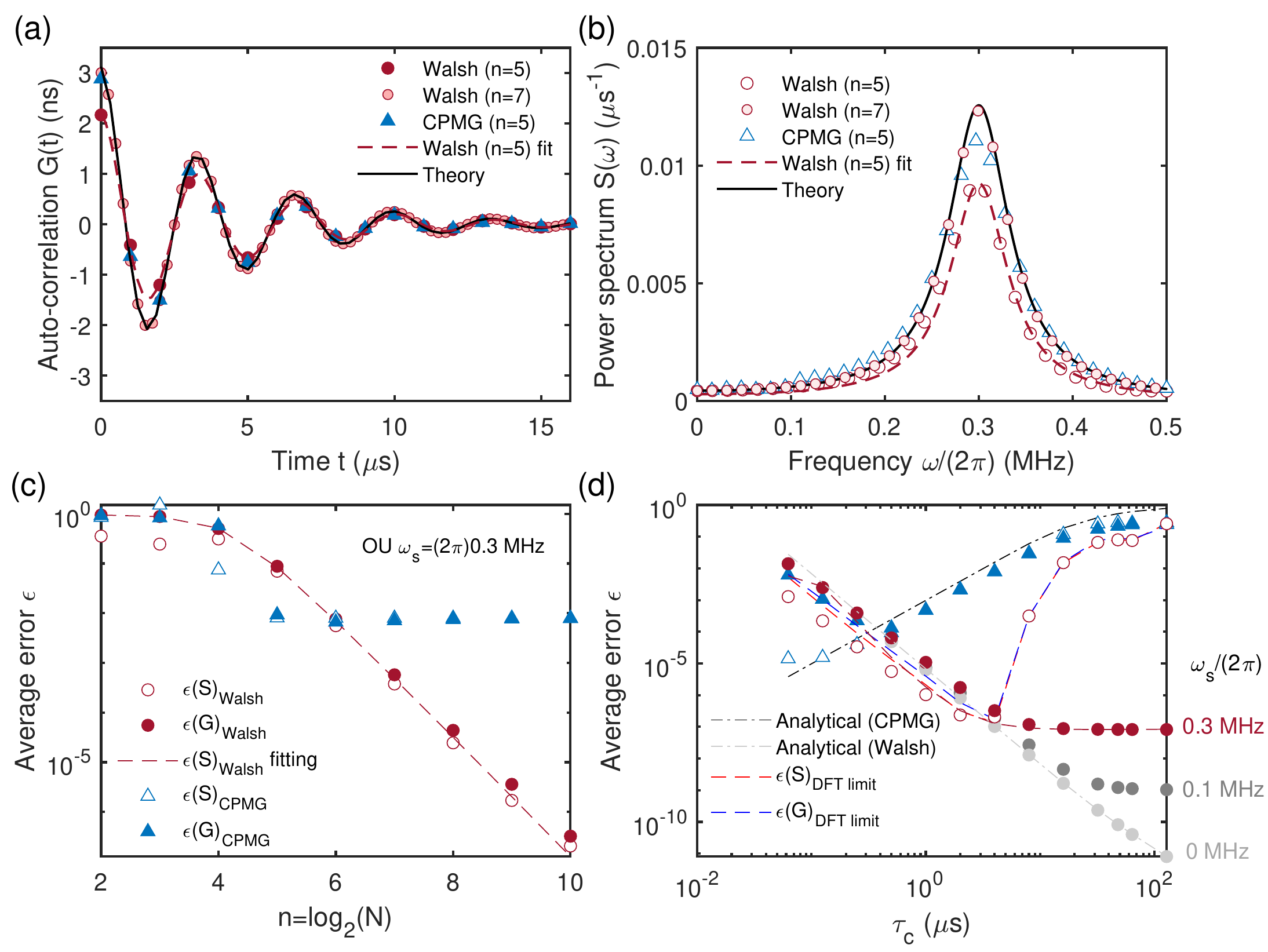}
\caption{\label{Fig_Comparison}
\textbf{Numerical results.}
(a)~Reconstructed auto-correlation function for simulated Ornstein-Uhlenbeck noise over a sampling time of $T=32~\mu\text{s}$ using Walsh QNS with $N=2^5=32$ (red disks) and $N=2^7=128$ (red shaded disks) sampling sequences \lv{vs.} CPMG QNS with $N=2^5=32$ sampling sequences. Data are shown up 16 $\mu$s to highlight the differences between the Walsh and CPMG reconstructed signals. The noise parameters are $\tau_c=4\ \mu$s, $b^2=0.003125\ \mu$s, $\omega_s=(2\pi)~0.3$~MHz.
(b)~Reconstructed power spectrum in the frequency domain using Walsh and CPMG. The ``Walsh fit" (dark red dashed curve) plots Eq.~\eqref{eq:S_OU_theory} with the fitting value of $b^2$, $\tau_c$, $\omega_s$ from the corresponding auto-correlation data in (a), and such a rule applies to other plots.
(c) Dependence of the average reconstruction error on the reconstruction order $n$. The OU noise parameters $\tau_c, b^2,\omega_s$ and sampling time $T$ are the same as (a). (d) Dependence of the average reconstruction error on the auto-correlation time $\tau_c$. The parameters $b^2$ and $T$ are the same as for (c), while $n=10$ is fixed and Walsh cases with $\omega_s=0,(2\pi)0.1$~MHz are also added as a comparison. Colors are the same as for (c), \Black{except for extra specifications}. The analytical curves (black and gray dashed curves) plot  theoretically predicted $\epsilon(S)$ and $\epsilon(G)$ in Eqs.~\eqref{eq:analyitical} and \eqref{eq:analytical_Walsh}. DFT-limited $\epsilon(S)$ (red) and $\epsilon(G)$ (blue) are the errors of the spectrum and auto-correlation obtained through DFT of theoretical $G(t_j)$ and $S(\omega_j)$, respectively.
}
\end{figure*}

\section{Operational performance} 

To demonstrate the power of Walsh QNS, we 
apply our method to reconstruct typical noise spectra and compare its performance -- defined in terms of the reconstruction error -- to conventional QNS protocols designed to use similar experimental resources. For concreteness, we numerically 
reconstruct the auto-correlation function of a fluctuating magnetic field sampled from an Ornstein-Unlenbeck (OU) noise model~\cite{uhlenbeck_theory_1930}~(Fig.~\ref{Fig_Comparison}), which is described by 
\begin{equation}
    G(t)=\langle\omega(t)\omega(0)\rangle=b^2e^{-\frac{|t|}{\tau_c}}\cos(\omega_st)
    \label{eq:G_OU_theory},
\end{equation}
where $b^2$, $\omega_s$, and $\tau_c$ characterize the strength, frequency, and correlation time of the noise, respectively~(Fig.\,\ref{Fig_Comparison}(b)). The associated frequency-domain power spectrum comprises two symmetric Lorentzian peaks:
\begin{equation}
    S(\omega)=\frac{b^2\tau_c}{1+(\omega-\omega_s)^2\tau_c^2}+\frac{b^2\tau_c}{1+(\omega+\omega_s)^2\tau_c^2}.
    \label{eq:S_OU_theory}
\end{equation}
The OU model was originally used to describe the dynamics of Brownian motion~\cite{uhlenbeck_theory_1930,gillespie_exact_1996,gillespie_mathematics_1996} and then found great success in describing the decoherence of a spin qubit dipolarly-coupled to a noisy spin bath~\cite{de_lange_universal_2010,hanson_coherent_2008,dobrovitski_decay_2009}. 

We benchmark the operational performance of Walsh QNS against a typical \lv{comb-based QNS protocol employing} CPMG sequences. To perform a fair comparison with the Walsh scheme with sampling time $T$  and order $N$, we design a set of CPMG sequences with a total experimental time $\sim NT$. In the CPMG scheme as shown in Fig.~\ref{Fig_WalshCPMG_main}(b)~\cite{SOM}, the zeroth sequence is a Ramsey sequence, \lv{whereas} for each $k\in\{1,\cdots,N/2\}$ the $2k-1$, $2k$ sequences have $2k$ $\pi$-pulses with intervals $\tau_{2k-1}=T/(2k-1)$, $\tau_{2k}=T/(2k)$, respectively. This protocol reconstructs the  noise spectrum by sampling $S(\omega_k)$ at equidistant frequencies,  $\omega_k=k\pi/T$, $k=0,1,\cdots,N$. The reconstruction is done  using the inverse of Eq.~\eqref{eq:CPMG_chi}, where higher harmonics effects are taken into account using methods in Refs.~\cite{yuge_measurement_2011,alvarez_measuring_2011}. \lv{Since for a classical process the noise spectrum is symmetric,} $S(\omega)=S(-\omega)$, we obtain the auto-correlation $G(t_j)$ with a sampling $t_j=\frac{jT}{N}$ ($j=0,\cdots,N$) similar to the Walsh method, by applying the inverse DFT~\cite{SOM}. Examples of the reconstruction of an OU noise at 0.3 MHz frequency are shown in Figs.~\ref{Fig_Comparison}(a,b), which show that the Walsh reconstruction improves significantly at larger $n$ ($n=7$) and outperforms the CPMG method. 

We quantify the operational performance of spectral reconstruction protocols with the normalized reconstruction error in both the time  and frequency domains,
\begin{equation}
    \epsilon(A)\equiv \frac{\sum_{i} [A(i)-A_0(i)]^2}{\sum_i A_0(i)^2},
\end{equation}
where $A_0$, $A$  are the theoretical and estimated values of the reconstructed function $A$, which is either $G$ or $S$.
Indeed for an OU noise, one can analytically calculate the reconstruction error. For the CPMG scheme limited by the error in the $\delta$-function approximation, a precise analysis in Ref.~\cite{szankowski_accuracy_2018} under $\tau_c\lesssim T$ yields 
\begin{equation}
\epsilon(S)\approx \Big(1-e^{-\frac{T}{\tau_c}}\Big)^2\frac{\tau_c^2}{T^2} ,
    \label{eq:analyitical}
\end{equation}
showing that the error grows as $\tau_c$ and can only be reduced by increasing $T$.   The Walsh scheme is limited by the error in approximating $G$ with the discretized $\overline{G}_N$. We then analytically obtain the reconstruction error for an OU noise at zero frequency when $\tau\lesssim\tau_c$,
\begin{equation}
    \epsilon(G)\approx\frac{\tau^3}{\tau_c^3}\frac19\left(1+\coth(\frac{T}{\tau_c})\right),   \label{eq:analytical_Walsh}
\end{equation}
showing that the error decreases not only as $\tau_c$ increases, but also as the reconstruction order $n$ increases, resulting in decreasing $\tau=T/2^{n}$. This feature makes the Walsh scheme distinct from the CPMG scheme, which can only suppress the reconstruction error by increasing the time $T$ (and not $n$ alone, see Appendix~\ref{App_ReconstructionErr} for more detail).

We numerically compute the reconstruction error as a function of the reconstruction order $n$ in Fig.~\ref{Fig_Comparison}(c). As $n$ increases, the average error decreases initially for both the Walsh and the CPMG methods; however, the error for CPMG soon saturates at large $n$, while the error for Walsh keeps decreasing with an exponential behavior. This clearly shows that the reconstruction error for Walsh QNS can be suppressed to arbitrarily small values as long as the number of sequences is large enough, as indeed its error is only limited by the sampling time-window. 
Instead, sampling with larger $n$ in  CPMG  only provides additional information in a higher-frequency range beyond the dominant feature of the target spectrum (concentrated at a low frequency), which does not improve the reconstruction of the overall spectrum. 

To further validate that the Walsh method outperforms the CPMG method under a large correlation time $\tau_c$, we compute the $\tau_c$ dependence of the average reconstruction error~(Fig.~\ref{Fig_Comparison}(d)). We exclude  potential errors introduced by the DFT 
by comparing the error for the CPMG spectrum to the error for the Walsh auto-correlation. The calculated CPMG error $\epsilon(S)$ matches the analytical prediction (black dashed curve) in Eq.~\eqref{eq:analyitical}, which increases with $\tau_c$. In contrast, the Walsh error $\epsilon(G)$ decreases with $\tau_c$ as predicted by Eq.~\eqref{eq:analytical_Walsh} (gray dashed curve).  
For the OU noise centered at non-zero frequencies, the error saturates at large $\tau_c$, which is due to the imperfect sampling of the oscillatory feature of the auto-correlation. Thus, under the same reconstruction sequences, a faster oscillation (larger $\omega_s$) in the OU model has a larger saturation error, due to  worse sampling -- as shown in the comparison between $\omega_s/(2\pi)=0,0.1, \text{and}~0.3$~MHz in  Fig.~\ref{Fig_Comparison}(d).

Despite providing an ``exact" reconstruction of the noise in the time domain, limited only by the sampling order, it is still necessary to  analyze the Walsh reconstruction error  in the frequency domain, which might be additionally limited by the error in DFT. 
Such a limitation may come both from a finite time range, which does not capture the full decay of the noise auto-correlation, or from finite sampling points, which do not capture the high-frequency oscillatory feature. In Fig.~\ref{Fig_Comparison}(d), we exclude the effect of insufficient sampling for the high-frequency feature by setting a large $N=2^{10}$, such that the limitation is only given by the finite time range. We plot the DFT of the theoretical $G(t_j)$ (red dashed curve), which dominates the error for the Walsh spectrum $\epsilon(S)$ at large $\tau_c$. Similarly, we plot the inverse DFT of the theoretical $S(\omega_j)$ (blue dashed curve), which dominates the error for the CPMG auto-correlation $\epsilon(G)$ at small $\tau_c$ due to the imperfect capture of high frequency components with a finite frequency range. We note that with the knowledge of the \lv{functional form of the} noise model, these limitations can be overcome by fitting and extrapolating the data to an infinite range before performing Fourier transform (dark red dashed curves).

\section{Experimental demonstration} 
\begin{figure}[htb]
\centering \includegraphics[width=0.48\textwidth]{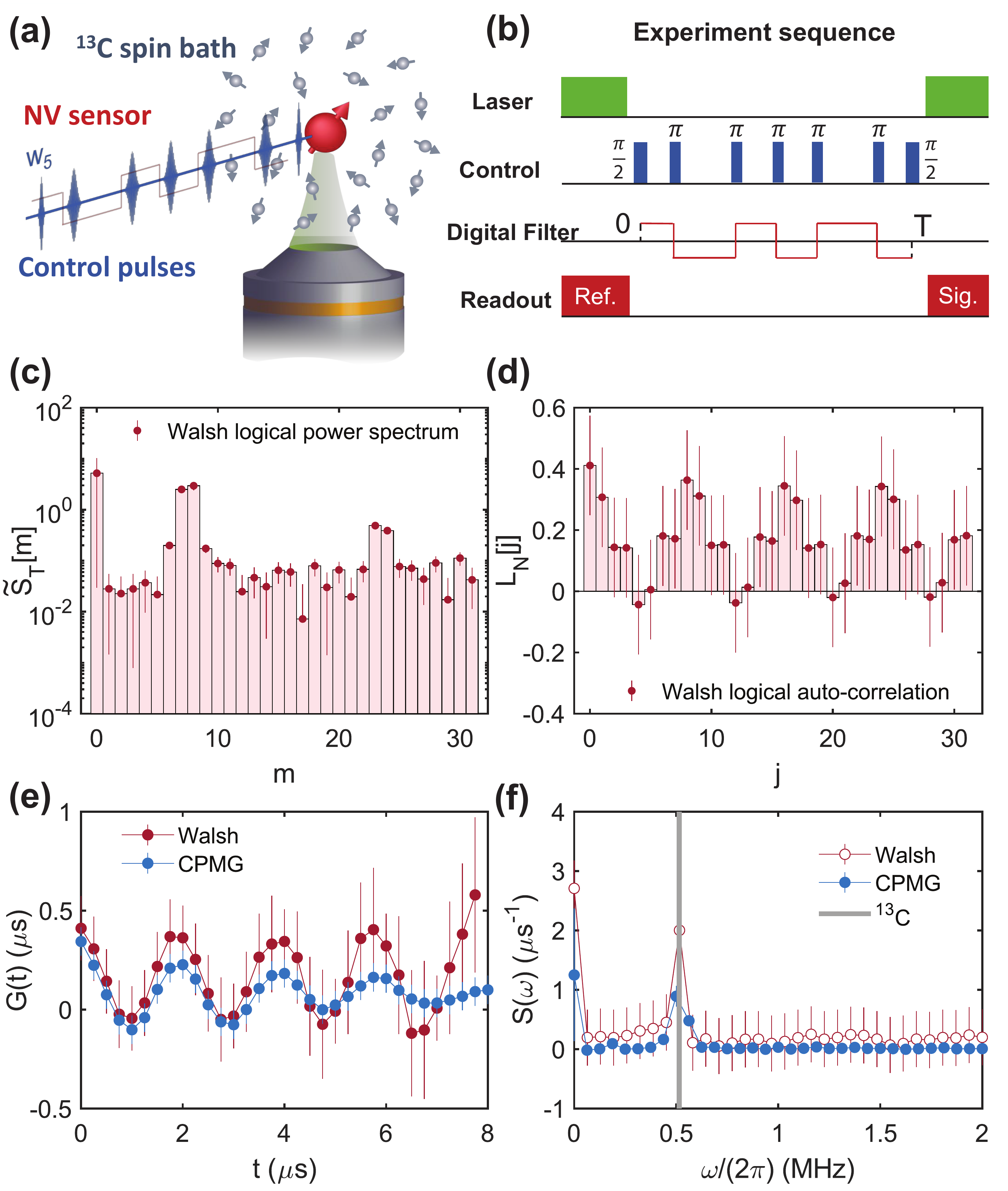}
\caption{\label{Fig_Experiment}
\textbf{Experimental results}.
(a)-(b)~A control sequence of five 
$\pi$-pulses~(blue) modulates the electronic spin of a single nitrogen-vacancy center in diamond to sample the fifth coefficient of the logical power spectrum of the noise generated by a bath of $^{13}$C nuclear spins. The control pulses are applied at the node of the fifth Walsh sequence to synthesize an effective digital filter (red). A green laser at $532~\text{nm}$ focused through a high-resolution microscope objective is used to optically polarize and readout the state of the electronic spin.
(c)~Measured logical power spectrum $\tilde{S}_T[m]$ in the sequency domain with $N=2^5=32$ Walsh sequences of duration $T=8\ \mu$s.
(d)~Reconstructed logical auto-correlation function, $\bar{L}_N[j]$.
(e)~Reconstructed auto-correlation function, $G(t)$, \lv{using Walsh and CPMG QNS}. The CPMG sequences have the same duration as the Walsh sequences.
(f)~Reconstructed power spectrum in the frequency domain, $S(\omega)$, \lv{using Walsh and CPMG QNS}. The vertical bar indicates the expected precession frequency of $^{13}$C nuclear spins.
}
\end{figure}

To demonstrate the applicability of our method in an operational setting, we performed optically-detected pulsed magnetic resonance experiments on the electronic spin of a single NV center in diamond at ambient temperature~\cite{wang_nanoscale_2021, liu_nanoscale_2019}~(Fig.~\ref{Fig_Experiment}(a)). We aligned a static magnetic field of $\sim$460~G along the NV axis to lift the degeneracy between the $|m_S=\pm1\rangle$ electronic ground states and optically polarized the $^{14}$N nuclear spin to the $\ket{m_I=+1}$ nuclear spin state via an excited-state level anti-crossing (ESLAC)~\cite{jacques_dynamic_2009}.
The NV center was located at a depth of $\approx30~\mu$m, where the spin bath is mainly composed of the $^{13}$C nuclear spins with a concentration $\leq1.1~\%$ and a nuclear Larmor frequency of $\approx0.5~\text{MHz}$. Although the nuclear spin bath is a quantum many-body system, its effective interaction with the electronic spin at such a large field over multiple experimental realizations can be \lv{accurately} described by an effective longitudinal fluctuating classical field~\cite{hernandez-gomez_noise_2018,reinhard_tuning_2012}.

We coherently modulated the spin qubit encoded in the $|m_S=0\rangle$ and $|m_S=-1\rangle$ states of the electronic spin using sequences of microwave pulses delivered through a straight copper wire. We sampled the logical power spectrum of the noise with up to $N=2^5=32$ Walsh and CPMG sequences~(Fig.~\ref{Fig_Experiment}(c)) of duration $T=8~\mu\text{s}$, from which we reconstructed estimates of the logical auto-correlation function, auto-correlation function, and power spectrum~(Figs.~\ref{Fig_Experiment}(d-f)). 
The auto-correlation function reconstructed using both Walsh and CPMG methods shows an oscillation at around $0.5$~MHz (Fig.~\ref{Fig_Experiment}(e)), which is observed as a peak in the power spectrum (Fig.~\ref{Fig_Experiment}(f)), matching the predicted Larmor frequency of the nuclear spin $^{13}$C. These results demonstrate the applicability of Walsh reconstruction in characterizing the noisy environment of quantum devices.

The operational performance of Walsh QNS, \lv{in particular in terms of} reducing error bars or suppressing the amplitude increase in the auto-correlation function, could be further improved by addressing three sources of uncertainty, \lv{as we address next}. 

The first source of uncertainty is \lv{bias arising from a finite statistical sample.} 
In Eqs.\,\eqref{eq:chi_theory} and~\eqref{eq:LN}, the arithmetic and logical auto-correlations are well-defined for sufficient averaging over noise realizations, such that $\langle\omega(t)\omega(0)\rangle=G(t)+\epsilon(t)\approx G(t)$, where the approximation error induced by $\epsilon(t)$ becomes non-negligible if the statistical ensemble of noise trajectories is too small. For an OU noise with a correlation time $\tau_c$, the relative variance of the auto-correlation function increases almost exponentially with $t$ due to the exponential decrease of its average value, which leads to larger relative error of the reconstructed auto-correlation at larger $t$. A reliable way to suppress the upper bound of this error is to increase the number of experimental averages, as we further discuss 
in the Supplemental Materials~\cite{SOM}.

The second source of uncertainty is the finite contrast $c$ of the interferometric signal ${\cal S}$, from which the coherence factor \lv{$\chi=-\log{(2{\cal S}-1)}$} 
is estimated with a relative error $\Delta\chi=\sigma_\chi/\chi=(2\sigma_P/c) e^{\chi}/\chi$, where $\sigma_ P$ is the uncertainty of the projective measurement, possibly due to photon shot-noise or spin-projection noise~\cite{barry_sensitivity_2020}. When $\sigma_P$ is independent of $\chi$, e.g., for a shot-noise limited measurement, then $\Delta\chi$ is minimum at $\chi=1$ and diverges at both small and large values of $\chi$. The sampling time should thus be chosen neither too long nor too short to achieve the optimal value of $\chi\sim 1$; however, at fixed sampling time, different sampling sequences have different $\Delta\chi$, and the large $\Delta\chi$ from a few coefficients contaminate the reconstruction -- resulting in larger error bars on the reconstructed spectrum, as well as larger error bars on $G(t)$ at longer times~(see Fig.~\ref{Fig_Experiment}(e))~\cite{SOM}. This error propagation issue can be solved by thresholding small coefficients with large $\Delta\chi$, performing a weighted reconstruction, or amplifiying the signal, e.g., by sampling the noise with $M$ concatenated sampling sequences and fitting the data to an exponential function of $M$~\cite{szankowski_accuracy_2018,szankowski_environmental_2017}.  

The third source of uncertainty stems from coherent modulation of the signal due to the \lv{fact that the actual noise environment the sensor experiences is more complex than the classical model used thus far accounts for. In particular, it includes strongly coupled proximal nuclear spins which act as a genuinely quantum noise source. In the presence of such a ``quantum-classical noise environment,'' the measured signal is modified to ${\cal S}(T)=(1+e^{-\chi(T)}\prod_j M^{(j)}(T))/2$, where $M^{(j)}(T)$ accounts for the influence of coherent oscillations due to individual proximal spins~\cite{hernandez-gomez_noise_2018,taminiau_detection_2012,kolkowitz_sensing_2012,SOM}. The collapse of the signal associated with spin-bath characteristic frequencies} might thus prevent directly extracting $\chi(T)$ from ${\cal S}(T)$ for certain Walsh sequences. Following typical CPMG methods, which extract both the quantum and classical noise contributions by fitting the time sweep data to a well-developed analytical formula, this problem could be solved by using larger sampling time $T$ and reconstruction order $N$ to suppress the contribution of the quantum bath~(see Appendix~\ref{App_QC} and Supplemental Materials~\cite{SOM} for further discussion). 

\section{Conclusions and outlook}

To conclude, we \lv{have proposed} a digital noise spectroscopy method based on Walsh sequences, which resolves the difficulty in implementing \lv{standard comb-based} reconstruction approaches that require non-linear deconvolution algorithms or delta-like filter-function approximations. By applying a sequence of linear transformations, our method achieves a direct reconstruction of the noise auto-correlation with an accuracy only limited by the sampling in the time domain. \lv{Thanks to the fact that Walsh spectroscopy may be understood within a more general frame-based approach \cite{Chalermpusitarak21}, it follows that the information inferred about the noise is provably sufficient to accurately predict the evolution of the system under {\em arbitrary} digital dephasing-preserving control (i.e., arbitrary digital sequences of $\pi$ pulses), beyond the Walsh sequences used in the spectroscopy protocol itself.} 
Our experimental demonstration using a single NV center in diamond further shows the applicability of our method to characterize the noise environment of realistic quantum devices. Although the experimental deviation from a classical noise model can be eliminated by various strategies, the same deviation can also be used as a sensitive probe of the dynamics of the non-classical environment, as discussed below.

Extending the Walsh approach to characterize complex quantum-classical environments is of significant interest for future research. Indeed, the presence of quantum (and/or non-Gaussian) components in the noise has been shown to give rise to novel phenomena, such as anomalous or distinctive decoherence behavior~\cite{huang_observation_2011,zhao_anomalous_2011,google_gate}. 
\lv{Our numerical results demonstrate how the sensitive response of the Walsh reconstruction technique to the quantum spin bath offers a}  potential advantage in reconstructing quantum-classical environments (see Appendix~\ref{App_QC}). It would be then interesting to systematically study how Walsh sequences modulate the dynamics of a quantum system subjected to such noise sources. This could lead to further methods exploiting Walsh sequences to reconstruct the higher-order spectra (``polyspectra'') of a stationary non-Gaussian noise process~\cite{norris_qubit_2016,sung2019}, or to perform noise cross-spectroscopy using multiple qubits~\cite{szankowski_spectroscopy_2016,paz-silva_multiqubit_2017,kwiatkowski_decoherence_2018} -- for instance, using coherent control on coupled systems of electron spins~\cite{cooper_2020, cooper_2019, degen_2021}. The direct access to the noise auto-correlation provided by Walsh spectroscopy as well as its optimal performance under complementary conditions than \lv{comb-based} frequency-domain spectroscopy methods would be highly beneficial to characterize a broad range of environmental noise sources.

\acknowledgements

It is a pleasure to thank Won Kyu Calvin Sun and Han Chen for useful discussions. L.V. is also especially grateful to Gerardo Paz-Silva for valuable input and collaborations over the years. This work was partly supported by  HRI-001835, NSF EECS1702716 and PHY1734011. \lv{Work at Dartmouth was supported in part by the U.S. Army Research Office through MURI Grant No. W911NF1810218}.

\begin{appendix}
\section{Theory}
\label{App_Theory}

In this section, we highlight the challenges of sampling and reconstructing continuous functions using digital filters, provide additional insights to better understand the connections between the logical and arithmetic domains, and formalize the construction of the shuffling transformation and its matrix representation. We present a complementary, albeit more formal, derivation of the results presented in the main text using concepts from the theory of integration. Detailed derivations using matrix representation are included in Supplemental Materials~\cite{SOM}.

We consider a single spin qubit interacting with a fluctuating field oriented along its quantization axis described by a Hamiltonian $\mathcal{H}(t)=\omega(t)\sigma_z/2$. We perform a sequence of Ramsey measurements while coherently modulating the state of the spin with a sequence of $m$ $\pi$-pulses, effectively generating the digital filter $f_m(t)$ over the time interval $[0, T]$, which we choose among the complete set of Walsh functions $\{f_m\}_{m=0}^{N-1}$ of order $N=2^n$. The Walsh functions are piecewise-continuous functions defined over $N$ contiguous intervals of length $\tau=T/N$, which can be represented as discrete functions, $f_m(t)=w_m(t/\tau)=W[m,j]$ for $t\in[j\tau, (j+1)\tau)$, $m,j\in[0, 1, \cdots, N-1]$, and $W[m,j]$ is the $[m,j]$ element of the Walsh matrix of size $N$.

The dynamical phase acquired by the quantum sensor (Eq.~\eqref{eq:phase}) is $\phi_m(T)=\hat{\omega}_m(T)\cdot T$,
where the effective precession frequency $\hat{\omega}_m$ is exactly the $m$-th Walsh coefficient of the longitudinal field,
\begin{eqnarray}
\hat{\omega}_m(T)&=&\frac{1}{T}\int_0^T \omega(t)f_m(t)dt.
\end{eqnarray}

Assuming the external time-varying field $\omega(t)$ to be the representative of a zero-mean ($\ave{\omega(t)}=0$), stationary, Gaussian stochastic process, which is fully described by its second 
\lv{cumulant} (variance), the mean normalized signal obtained after averaging over a series of $M$ sequential measurements is
\begin{eqnarray}
\mathcal{S}_m(T)&=&\ave{\exp(-i\phi_m(T))}=\exp(-\chi_m(T)).
\end{eqnarray} 
Performing a cumulant expansion, the \emph{decoherence exponent} is given by the variance of the Walsh coefficient, $\chi_m(T)=\ave{\hat{\omega}_m^2(T)}T^2/2$, which can be expressed in terms of the auto-correlation function of the stochastic process $G(t_1, t_2) $, see Eq.~\eqref{eq:chi_theory}. We use the assumption of stationarity and the composition property of Walsh functions under multiplication, $f_m(t_1)f_m(t_2)=w_m(\lfloor \frac{t_1}{\tau}\rfloor \oplus \lfloor \frac{t_2}{\tau}\rfloor)$ \lv{(where $\oplus$ denotes the binary addition),} to obtain
\begin{align}
\chi_m(T)&=\frac{1}{2}\iint G(t_1-t_2)w_m\Big(\lfloor \frac{t_1}{\tau}\rfloor \oplus \lfloor \frac{t_2}{\tau}\rfloor\Big) dt_1dt_2\label{eq:sRw}.
\end{align}

Applying the change of variables, $u(t_1,t_2)=t_1-t_2\in(-T,T)$ and $v(t_1,t_2)=\lfloor \frac{t_1}{\tau}\rfloor \oplus \lfloor \frac{t_2}{\tau}\rfloor\in[0,N)$, and \lv{letting $d\mu(u,v)$ denote} the integration measure, Eq.~\eqref{eq:sRw} can be rewritten as
\begin{eqnarray}
\chi_m(T)&=&\frac{1}{2}\iint G(u)w_m(v)d\mu(u,v)\label{eq:sRu_wv}.
\end{eqnarray}
This equation clearly highlights the core difficulty in reconstructing $G(u)$ from measurements of $\chi_m(T)$, e.g., via direct inversion: the decoherence exponent is equal to an overlap integral between two functions that are defined over different domains -- the auto-correlation function over the \emph{arithmetic} domain and the Walsh function over the \emph{logical} domain. 

This apparent difficulty is elegantly lifted by introducing the shuffling transformation $T(u,v)$, which admits an explicit matrix representation derived below, connecting the arithmetic domain to the logical domain. Choosing $d\mu(u,v)=T(u,v)dudv$, we obtain
\begin{eqnarray}
\chi_m(T)&=&\frac{1}{2}\iint du G(u)T(u,v)w_m(v)dv\label{eq:RTw},
\end{eqnarray}
which has two equivalent representations depending on whether the integration is carried over the arithmetic domain ($u$) or the logical domain ($v$).

On the one hand, carrying the integration over the logical domain ($v$) brings us back to the conventional representation in the frequency domain,
\begin{eqnarray}
\chi_m(T)&=&\frac{1}{2}\int du G(u)\tilde{w}_m(u)\\
~&=&\frac{1}{2}\int d\omega S(\omega) \tilde{w}_m(\omega),
\end{eqnarray}
where 
\begin{equation}
\tilde{w}_m(u)\equiv\int T(u,v)w_m(v)dv=(w_m\ast\bar{w}_m)(u)
\label{eq:sFF}
\end{equation}
is the {\em shuffled FF} expressed in the arithmetic time domain, while 
$$\tilde{w}_m(\omega)\equiv\mathcal{F}\{\tilde{w}_m(u)\}=|\mathcal{F}\{w_m(v)\}|^2$$ is the frequency FF obtained by computing its (finite time) Fourier transform, $\mathcal{F}\{\cdot\}$.  Accordingly, the decoherence exponent is proportional to the overlap integral between the frequency filter and the power spectrum, $S(\omega)=\int_{-\infty}^{\infty}e^{-i\omega u}G(u)du$. 

It is worth noting that the Walsh frequency FFs have exact analytical expressions given by the product of trigonometric functions~\cite{robinson_1974}. \lv{However,}
reconstructing $S(\omega)$ from measurements of $\chi_m(T)$ with a set of digital filters whose frequency-filter representation is given by some non-trivial trigonometric functions is a hard non-linear inversion problem that can only be solved numerically, e.g., using deconvolution algorithms.

To avoid this non-trivial inversion problem, a common strategy is to use periodic [CPMG ($m=2^n$) or PDD ($m=2^n-1$)] sequences,  a subset of the Walsh basis that allow a simple  approximation when sampling the power spectrum with a large number of pulses at high-frequency.   Their  frequency FF can be then approximated by a $\delta$-like filter function, e.g., $\tilde{w}_{\text{CPMG}}{(\omega)}\approx\delta(\omega-\omega_\text{CPMG})$, such that $\chi_m(T)\approx\frac{1}{2}\int d\omega S(\omega) \delta(\omega-\omega_\text{CPMG})=S(\omega_{CPMG})/2$. Although powerful for solving certain operational problems and providing a simple intuitive picture, this method lacks in precision and generality.

Returning to the shuffling transformation, we note that, 
given Eq.\,\eqref{eq:sFF}, 
$T(u,v)$ can be explicitly computed (up to order $N$) by computing the inverse Walsh transform (of order $N$) of the convolution product of ${w}_m(t)$ and its time-reversed self $\bar{w}_m(t)=w_m(T-t)$, that is,  
$$T(u,v)=\mathcal{W}^{-1}\{\tilde{w}_m(u)\}=\frac{1}{N}\sum_{m=0}^{N-1} (w_m\ast\bar{w}_m)(u) w_m(v).$$ 
In turn, the pseudo-inverse of $T(u,v)$, $T^{-1}(v,u)$, can be retrieved from the normalization condition, $\iint T(u,v)T^{-1}(u,v) du \,dv/\iint du \,dv=1$. 

Back to Eq\,\eqref{eq:RTw}, carrying the integration over the arithmetic domain ($u$)instead  gives us
\begin{eqnarray}
\chi_m(T)&=&\frac{1}{2}\iint du G(u)T(u,v)w_m(v)dv\nonumber\\
~&=&\frac{1}{2}\int L(v)w_m(v)dv\label{eq:sSm}\\
~&\equiv&\tilde{S}_T(m),\nonumber
\end{eqnarray}
where the decoherence exponent is shown to be exactly the (finite-time) logical power spectrum evaluated over the finite interval $[0,T)$. The key result of this paper is contained in Eq.\,\eqref{eq:sSm}: the decoherence exponent is exactly the logical power spectrum evaluated in the sequency domain, which in turn is the Walsh transform of the logical auto-correlation function $L(v)$. 
This equivalence  between the decohence exponent and the logical power spectrum leads to a simple sampling and reconstruction protocol to estimate the auto-correlation function of a fluctuating field via coherent modulation of a single-qubit sensor.

\section{{Relation to frame formalism}}
\label{app:frames}

\lv{Within classical signal processing, the formalism of frames plays an instrumental role in a variety of applications where flexibility in representing a signal of interest is desirable or necessary \cite{frames_intro}. Being a digital basis over the interval $[0,T]$, the Walsh functions form a Parseval, ``self-dual'' discrete frame. That is, $\mathscr{F}\equiv \{ \phi_n\} = \{ \tilde{\phi}_n\} \equiv \tilde{\mathscr{F}} = \{ w_m\}_{m=0}^{N-1}$. Within the general frame-based approach to noise characterization introduced in Ref.\,\cite{Chalermpusitarak21}, Walsh-based reconstruction protocols were introduced and (numerically) shown to achieve QNS beyond the frequency domain -- including possibly {non-stationary} noise processes of both classical and quantum (non-commuting) nature.}

\lv{In order to establish contact with the present setting, note that several simplifications of the general formalism of \cite{Chalermpusitarak21} are possible. Specifically: 
\begin{itemize}
    \item Since the relevant noise is purely dephasing and classical, and the control is instantaneous and dephasing-preserving ($\pi$ pulses only), it suffices to consider the $z$-direction of the control matrix -- that is, the switching function $y_{z,z}(t)\equiv y_z(t)$. 
    \item Since the applied control is limited to digital $\pi$-pulse  sequences, we have $\mathscr{C}=\{ w_m\}_{m=0}^{N-1}$, where $m$ is the sequency. Thus, the Walsh sequences are not only the relevant frame, $\mathscr{F}$, but they also provide the {\em exact} control used to implement QNS. 
    \item Since, as noted, the Walsh functions are a basis, for any finite $N$ they satisfy exactly the ``finite-size frame condition'' [$\varepsilon=0$ in Eq.\,(13) of Ref.\,\cite{Chalermpusitarak21}].
\end{itemize}}

\lv{A natural starting point is provided by the overlap integral that determines the qubit decay in Eq.\,\eqref{eq:chi_theory}, namely, 
\begin{align}
\chi_m(T)\!=\!\frac12\int_0^T\!\!\!\int_0^T\!\!\!\!  y_z(t_1) y_z(t_2)\, \langle\omega(t_1)\omega(t_2)\rangle dt_1 dt_2.
\label{eq:chi_theory2}
\end{align}
On the one hand, by expanding each of the above switching functions in the Walsh basis,  
\[ y_z(t) 
= \sum_n
\bigg[ \int_0^T \! y_z(s) w_n(s) ds \bigg] w_n(t), \]
and substituting in Eq.\,\eqref{eq:chi_theory2}, we may collect the remaining factors into an equivalent of the usual noise spectrum in the frame language: 
\begin{align*}
\chi_m(T)& = \frac{1}{2} \sum_{n_1,n_2} \prod_{j=1,2} \bigg[ \int_0^T y_z(s_j) w_{n_j}(s_j) ds_j\bigg]  \\
& \quad \times\int_0^T\!\!\!\int_0^T  \langle\omega(t_1)\omega(t_2)\rangle w_{n_1}(t_1) w_{n_2}(t_2) dt_1 dt_2 \nonumber \\
& \equiv \sum_{n_1,n_2} \bigg [ \prod_{j=1,2} \int_0^T y_z(s_j) w_{n_j}(s_j) ds_j\bigg] \bar{S} (n_1,n_2), \nonumber 
\end{align*}
where in the last line we have introduced the {\em frame-based control-adapted} noise spectrum \cite{Chalermpusitarak21}, 
\[ \bar{S} (\vec{n}) \equiv \int_0^T \!dt_1 \!\int_0^{t_1}dt_2 \,
\langle\omega(t_1)\omega(t_2)\rangle \, w_{n_1}(t_1) w_{n_2}(t_2). \]
As long as the control modulation is effected by a Walsh sequence, we have $y_z(t)=w_m(t)$, and the relevant frame-based fundamental FF takes an especially simple form:
\begin{align*}
F^{(1)}_z(n, T) \equiv \int_0^T y_z(t) w_n(t) dt 
= T \,\delta_{m,n}. 
\end{align*}}

\lv{On the other hand, we could equally well expand each of the noise variables in the Walsh frame, 
\[ \omega(t) = 
\sum_n \!
\bigg[ \int_0^T \! \omega (s) w_n(s) ds \bigg] w_n(t). \]
By substituting again in Eq.\,\eqref{eq:chi_theory2}, rewriting the noise auto-correlation in terms of frames, and collecting the remaining factors into a FF, one now recovers a discrete, frame-based version of the usual frequency representation. Specifically, we have: 
\begin{align*}
\chi_m(T)& = \frac{1}{2} \! 
\sum_{n_1,n_2} \!\bigg[ \int_0^T \!\!\!\int_0^T\!\prod_j y_z(t_j)w_{n_j}(t_j) dt_1 dt_2\bigg] {S} (\vec{n}), 
\end{align*}
where the {\em frame-based standard-picture} second-order FF is given by
\[F_z^{(2)} (\vec{n},T)\equiv \int_0^T \!\!d t_1 \!\int_0^{t_2} \!\!dt_2 \,y_z(t_1)w_{n_1}(t_1) y_z(t_2)w_{n_2}(t_2),\]
while the corresponding frame-based standard-picture power spectrum reads
\[ {S} (\vec{n}) \equiv \int_0^T \!dt_1 \!\int_0^T dt_2 \,
\langle\omega(t_1)\omega(t_2)\rangle \, w_{n_1}(t_1) w_{n_2}(t_2).\]}

\lv{From the above expression, it immediately follows that 
\begin{equation}
\chi_m(T) = T^2 \bar{S}(m, m) =  \frac{1}{2} T^2 {S}(m,m). 
\end{equation}
Comparing with Eq.\,\eqref{eq:sSm}, one thus sees that the logical power spectrum in the sequency domain, $\tilde{S}_T(m)$, may be related to the {\em diagonal} elements of both the standard and the control-adapted noise spectrum, in this setting. } 

\lv{In general, however, since the appropriate frame $\mathscr{F}$ is chosen in such a way that {\em all} possible switching functions $\{y_{z,v}(t)\}$ supported by $\mathscr{C}$ are represented efficiently (with small error $\varepsilon$), it is the finite set of control-adapted spectra that enables for a ``model-reduced'' description and efficient characterization of the open system dynamics~\cite{Chalermpusitarak21}. As detailed in Appendix \ref{App_Theory}, the Walsh QNS protocol we have presented in this work infers the target auto-correlation function from the decoherence exponent by leveraging stationarity and transformations between different domains, using solely $\pi$-pulse control. In contrast, the Walsh QNS protocols introduced and numerically demonstrated in \cite{Chalermpusitarak21} are applicable to arbitrary digital control (not necessarily dephasing-preserving) as well as non-stationary noise environments, at the cost of requiring the use of additional control sequences, which include non-$\pi$ rotation angles. In this way, switching functions $y_{z,v}(t)$, $v\in \{x,y,z\}$, corresponding to sums and differences of Walsh functions may also be realized, and {\em non-diagonal} elements $\bar{S}(m, m')$ can be estimated from the dynamics as well. The digitized noise auto-correlation may then be obtained via the relationship 
\[G(t_1,t_2)= \sum_{m,m'} [\bar{S}(m,m')+\bar{S}(m',m)]\,w_m(t_1)w_{m'}(t_2). \]
}

\section{Reconstruction error}
\label{App_ReconstructionErr}

The main error source in Walsh reconstruction comes from approximating the discretized auto-correlation with the piece-wise average of the continuous one,
\begin{equation}
    G((j-k)\tau)\approx\overline{G}_N[j-k].
\end{equation}
By considering an OU noise model with $\omega_s=0$, we can  analytically compute the error in the reconstructed auto-correlation. We consider separately the following two cases: when $j>k$,
\begin{align}
    \overline{G}_N[j-k]&=\frac{N^2}{T^2}\int_{\frac{kT}{N}}^{\frac{(k+1)T}{N}}\!\!\int_{\frac{jT}{N}}^{\frac{(j+1)T}{N}}\!\!\!\!dt_1dt_2G(t_1-t_2)
    \nonumber\\&=-\frac{1}{\tau^2}b^2\tau_c^2 e^{-\frac{(j-k)\tau}{\tau_c}}(1-e^{-\frac{\tau}{\tau_c}})(1-e^{\frac{\tau}{\tau_c}})\nonumber\\
    &=G((j-k)\tau)\frac{\tau_c^2}{\tau^2}\Big(e^{\frac{\tau}{\tau_c}}+e^{-\frac{\tau}{\tau_c}}-2\Big),
\end{align}
and when $j=k$, 
\begin{align}
    \overline{G}_N[0]&=\frac{2}{\tau^2}\int_{\frac{jT}{N}}^{\frac{(j+1)T}{N}} dt_1\int_{\frac{jT}{N}}^{t_1} dt_2 b^2e^{-\frac{t_1-t_2}{\tau_c}}\nonumber\\
    &=\frac{2\tau_c}{\tau^2}b^2 \int_{\frac{jT}{N}}^{\frac{(j+1)T}{N}} dt_1 (1-e^{\frac{t-j\tau}{\tau_c}})\nonumber\\
    &=G(0)\frac{2\tau_c^2}{\tau^2} \Big(\frac{\tau}{\tau_c}-1+e^{-\frac{\tau}{\tau_c}}\Big).\nonumber
\end{align}

In both cases, the reconstruction error can be suppressed provided $\tau/\tau_c\ll1$, as evident  when expressing the $\overline{G}_N$ for both cases to its leading order in $\tau/\tau_c$,
\begin{align}
    \overline{G}_N[i]&\approx G(i\tau)\Big(1+\frac{1}{12}\frac{\tau^2}{\tau_c^2}\Big),\quad i\neq0,\\
    \overline{G}_N[0]&\approx G(0)\Big(1-\frac{1}{3}\frac{\tau}{\tau_c}\Big),\quad i=0.
\end{align}
Thus, the average reconstruction error of the auto-correlation is  dominated by its first point, 
\begin{align}
    \epsilon&=\frac{\sum_{i=0}^{N-1}[\overline{G}_N[i]-G(i\tau)]^2}{\sum_{i=0}^{N-1} G(i\tau)^2}\nonumber\\&\approx \frac{(\frac{\tau}{3\tau_c})^2G(0)^2+(\frac{\tau^2}{12\tau_c^2})^2\sum_{i=1}^{N-1}G(i\tau)^2}{G(0)^2+\sum_{i=1}^{N-1}G(i\tau)^2}\nonumber\\
    &\approx\frac{\tau^2}{9\tau_c^2}\frac{(1-e^{-2\tau/\tau_c})}{(1-e^{-2N\tau/\tau_c})}\\
    &\approx\frac{\tau^2}{9\tau_c^2}\frac{2\tau}{\tau_c}\frac{1}{(1-e^{-2N\tau/\tau_c})}\\
    &\approx\frac{\tau^3}{\tau_c^3}\frac19\bigg(1+\coth(\frac{T}{\tau_c})\bigg).
\end{align}
where \lv{as before $T=\sum_k \tau_k$ and} we have used the summation  $\sum_{i=0}^{N-1}G(i\tau)\lv{^2}=G(0)\lv{^2}[1+e^{-2\tau/\tau_c}+\cdots+e^{-2(N-1)\tau/\tau_c}]=G(0)\lv{^2}(1-e^{-2N\tau/\tau_c})/(1-e^{-2\tau/\tau_c})$.

\section{Quantum-classical bath}
\label{App_QC}
\begin{figure}[htbp]
\centering \includegraphics[width=0.495\textwidth]{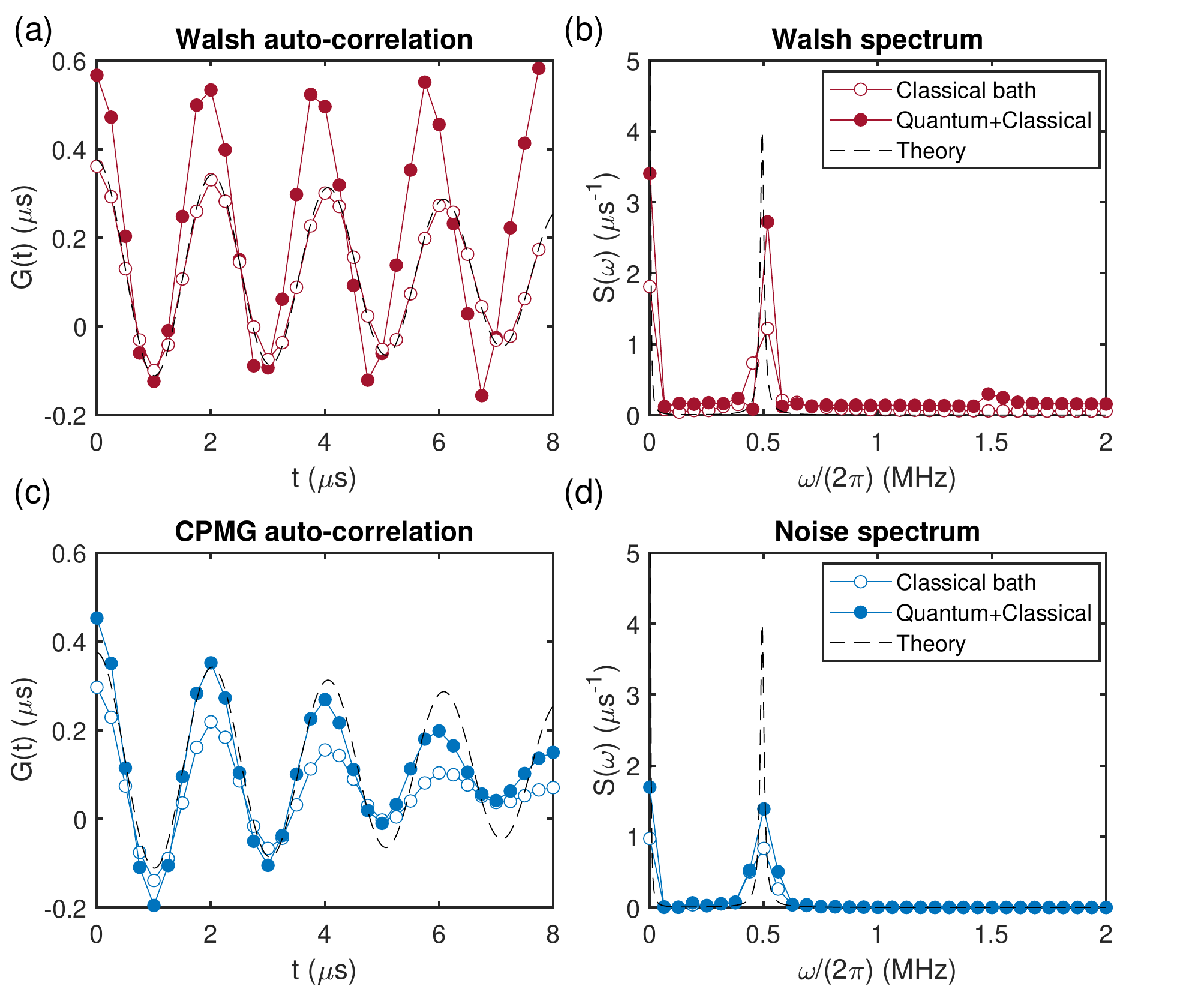}
\caption{\label{Supp_Fig_QuantumClassical} Simulation of a quantum-classical model. The sequence parameters are $T=8\ \mu$s, $N=32$. (a)-(b) Walsh reconstruction of the auto-correlation and noise spectrum. \lv{The quantum-classical noise model comprises a} nuclear spin with $\omega_L=-(2\pi)0.4926$~MHz, $A_\parallel=(2\pi)0.052$~MHz, $A_\perp=(2\pi)0.096$~MHz, as well as a classical noise bath with two OU components, with $b^2_{0,1}=0.125,0.25\ \mu$s, $\tau_{c0,c1}=80,16\ \mu$s at frequencies \lv{$\omega_s=0,\omega_L$}, respectively. For comparison, a classical bath without the quantum nuclear spin is also simulated and plotted. The theoretical curves for the classical noise are given by $G(t)=b_0^2e^{-t/\tau_{c0}}+b_1^2e^{-t/\tau_{c1}}\cos(\omega_L t)$ and $S(\omega)=2{b_0^2\tau_{c0}}/(1+\omega^2\tau_c^2)+{b_1^2\tau_{c1}}/(1+(\omega-\omega_L)^2\tau_c^2)+{b_1^2\tau_{c1}}/({1+(\omega+\omega_L)^2\tau_{c1}^2})$ are shown in dashed lines. (c)-(d) CPMG reconstruction of the auto-correlation and noise spectrum under the same noise model. }
\end{figure}

Although our theoretical analysis and experimental demonstration are based on the assumption that the \lv{noise enviroment coupling to the qubit sensor may be treated as a classical Gaussian process,
in a more general case a model comprising both a 
few nearby quantum spins and a far-away classical bath is appropriate}~\cite{kolkowitz_sensing_2012,hernandez-gomez_noise_2018,childress_coherent_2006,taminiau_detection_2012,reinhard_tuning_2012}. Here, we briefly discuss and compare the error introduced by the quantum spin \lv{component} when reconstructing the classical part for both Walsh and CPMG methods, \lv{still working under the assumptions of stationarity and Gaussianity, so that noise effects are fully accounted for by the two-point noise correlation function $G(t)$}.

We assume that the sensor couples to a few nearby nuclear spins through hyperfine interactions. The evolution of each nuclear spin ($j$) is dependent on the state of the sensor spin and the Hamiltonian can be further written as $ H=\sum_j \ket{0}\!\bra{0}\otimes H_{0}^{(j)}+\ket{1}\!\bra{1}\otimes H_{1}^{(j)}$, where  the intra-bath interaction of different nuclear spins are neglected, such that their evolutions are separable.
The initial state of the sensor is prepared to $\ket{+}$, \lv{while  the nuclear spins may be taken to be fully mixed}, thus the initial joint density matrix is $    \rho_0=\ket{+}\!\bra{+}\otimes \frac12 \mathds{1}^{(1)}\otimes \frac12 \mathds{1}^{(2)}\otimes\cdots$.
\lv{Accordingly, the sensor-plus-spins state at time $T$ is determined by $\rho(t)=U(T)\rho_0U^\dagger(T)$,
where} $U(T)=\mathcal{T}\exp(-i\int_0^T H(t)dt).$ To deal with such an operator, one can either seek a perturbative expansion  through the Magnus expansion~\cite{hernandez-gomez_noise_2018}, or continue to analyze the spin-dependent evolution~\cite{taminiau_detection_2012}. Here we follow the latter method and separate the time from $0$ to $T$ into $N+1$ \lv{segments} with duration $\tau_1,\tau_2,\tau_3,\cdots,\tau_{N+1}$, \lv{corresponding to the application of} $N$ instantaneous spin-flip pulses. Then the evolution operator $U(T)$ can be expressed as
\begin{equation}
    U(T)= \begin{cases}
\ket{0}\!\bra{0}\otimes U_0+\ket{1}\!\bra{1}\otimes U_1 & \text{$N$ even}, \\
\ket{1}\!\bra{0}\otimes U_0+\ket{0}\!\bra{1}\otimes U_1 & \text{$N$ odd}, \\
\end{cases}
\end{equation}
where the conditional evolution operators acting on the nuclear spins are 
\begin{align}
    U_{0,1}&=U_{0,1}^{(1)}\otimes U_{0,1}^{(2)}\otimes\cdots
\end{align}
with 
\begin{align}
    U_0^{(j)}=\left[e^{-iH_{0(1)}^{(j)}\tau_{N+1}}\cdots e^{-iH_{1}^{(j)}\tau_2}e^{-iH_{0}^{(j)}\tau_1}\right],\\ U_1^{(j)}=\left[e^{-iH_{1(0)}^{(j)}\tau_{N+1}}\cdots e^{-iH_{0}^{(j)}\tau_2}e^{-iH_{1}^{(j)}\tau_1}\right].
\end{align}
When also taking the classical noise bath into account, the final population in $\ket{+}$ at time $T$ is measured as
\begin{equation}
    \lv{{\cal S}(T)}=\frac12\bigg[1+e^{-\chi(T)}\Re\bigg(\prod_j M^{(j)}(T)\bigg)\bigg],
    \label{Supp_eq:Signal}
\end{equation}
where
\begin{equation}
    M^{(j)}(T)=\frac12\Tr^{(j)}[U_0^{(j)}(T)U_1^{(j)\dagger} (T)]
\end{equation}
corresponds to the contribution of \lv{the $j$-th} quantum nuclear spin, and $\chi(T)$ is due to the classical bath.

The error in  the noise reconstruction assuming a pure classical model comes from still using the classical signal formula ${\cal S}=(1+e^{-\chi})/2$, instead of the corrected one in Eq.\,\eqref{Supp_eq:Signal}.
To mimic the experimental results, we simulate the noise reconstruction under the same experimental condition in Fig.~\ref{Fig_Experiment} in the main text. We use a noise model composed of a single quantum nuclear spin as well as a classical spin bath that includes a zero-frequency OU noise source and a shifted OU noise with the Larmor frequency of the $^{13}$C nuclear spin. In Fig.~\ref{Supp_Fig_QuantumClassical}, we compare the reconstruction with and without the existence of the quantum spin. The results show that, for the qubit sensor subjected to a quantum-classical noise bath, the Walsh reconstructions present a larger deviation when still assuming the noise is classical, while the CPMG reconstructions \lv{are comparatively} less affected. 

\end{appendix}
\bibliography{main_text} 

\clearpage
\pagebreak
\widetext
\setcounter{section}{0}
\setcounter{equation}{0}
\setcounter{figure}{0}
\setcounter{table}{0}
\setcounter{page}{1}
\makeatletter
\renewcommand{\theequation}{S\arabic{equation}}
\renewcommand{\thefigure}{S\arabic{figure}}
\renewcommand{\thesection}{S\arabic{section}}


\begin{CJK*}{UTF8}{}
\title{Supplemental Materials: \\
Digital noise spectroscopy with a quantum sensor}

\author{Guoqing Wang \CJKfamily{gbsn}(王国庆)}\email[]{gq\_wang@mit.edu}
\affiliation{
   Research Laboratory of Electronics, Massachusetts Institute of Technology, Cambridge, MA 02139, USA}
\affiliation{
   Department of Nuclear Science and Engineering, Massachusetts Institute of Technology, Cambridge, MA 02139, USA}

\author{Yuan Zhu}
\affiliation{
   Research Laboratory of Electronics, Massachusetts Institute of Technology, Cambridge, MA 02139, USA}
\affiliation{
   Department of Nuclear Science and Engineering, Massachusetts Institute of Technology, Cambridge, MA 02139, USA}
\affiliation{
   Department of Electrical Engineering and Computer Science, Massachusetts Institute of Technology, Cambridge, MA 02139, USA}

\author{Boning Li}
\affiliation{
   Research Laboratory of Electronics, Massachusetts Institute of Technology, Cambridge, MA 02139, USA}
\affiliation{Department of Physics, Massachusetts Institute of Technology, Cambridge, MA 02139, USA}
   
\author{\mbox{Changhao Li}}
\thanks{Current address: Global Technology Applied Research, JPMorgan Chase, New York, NY 10017 USA}
\affiliation{
   Research Laboratory of Electronics, Massachusetts Institute of Technology, Cambridge, MA 02139, USA}
\affiliation{
   Department of Nuclear Science and Engineering, Massachusetts Institute of Technology, Cambridge, MA 02139, USA}

\author{Lorenza Viola}
\affiliation{\mbox{Department of Physics and Astronomy, Dartmouth College, 6127 Wilder Laboratory, Hanover, NH 03755, USA}}

\author{Alexandre Cooper}
\affiliation{Institute for Quantum Computing, University of Waterloo, Waterloo, ON N2L 3G1, Canada}

\author{Paola Cappellaro}\email[]{pcappell@mit.edu}
\affiliation{
   Research Laboratory of Electronics, Massachusetts Institute of Technology, Cambridge, MA 02139, USA}
\affiliation{
   Department of Nuclear Science and Engineering, Massachusetts Institute of Technology, Cambridge, MA 02139, USA}
\affiliation{Department of Physics, Massachusetts Institute of Technology, Cambridge, MA 02139, USA}

\maketitle

\end{CJK*}	

\begin{widetext}
\tableofcontents
\section{Dynamical-decoupling based noise spectroscopy methods}
\subsection{Decoherence under stationary Gaussian dephasing noise}

We discuss the dephasing of a qubit initial state $\ket{+}=(\ket{0}+\ket{1})/\sqrt{2}$ subject to a time-dependent noise field $\omega(t)\sz/2$ and to a sequence of instantaneous $\pi$ pulses, corresponding to a modulation function $f(t)$. The accumulated phase is then
\begin{equation}
    \varphi(T)=\int_0^T\omega(t)f(t)dt.
\end{equation}
The measured signal under sufficient statistical averaging is
\begin{equation}
    S(T)=\frac{1}{2}(1+\langle\cos(\varphi(T))\rangle)=\frac12(1+e^{-\chi})=\frac12(1+e^{-\langle\varphi^2\rangle/2}),
\end{equation}
where we assume a stationary and zero-mean noise such that
\begin{equation}
    \chi=\frac12\langle\varphi^2\rangle=\frac12\int_0^T\int_0^T dt_1 dt_2 \langle\omega(t_1)\omega(t_2)\rangle f(t_1)f(t_2).
    \label{Supp_eq:chi_theory}
\end{equation}
We define the auto-correlation of the noise as 
\begin{equation}
    G(t_1,t_2)=\langle\omega(t_1)\omega(t_2)\rangle.
    \label{Supp_eq:G_theory}
\end{equation}
For stationary and zero-mean noise, the above correlation only depends on the difference between two times, such that $G(t_1,t_2)=G(|t_1-t_2|)$.

The attenuation function $\chi$ can be naturally expressed as an integral in frequency space, 
\begin{align}
    \chi&=\frac12\int_0^T\int_0^T dt_1 dt_2 f(t_1)f(t_2) \int_{-\infty}^{+\infty}\frac{d\omega}{2\pi}S(\omega)e^{i\omega (t_1-t_2)}\\&=\frac12\int_{-\infty}^{+\infty}\frac{d\omega}{2\pi}S(\omega)[\int_0^Tf(t_1)e^{i\omega t_1}dt_1][\int_0^Tf(t_2)e^{-i\omega t_2}dt_2]\\&=\frac12\int_{-\infty}^{+\infty}\frac{d\omega}{2\pi}S(\omega)|F(\omega)|^2
\end{align}
where we have defined a filter function 
$F(\omega)=|\int_0^Tf(t)e^{i\omega t}dt|$.

The attenuation factor can also be calculated in time space as in Ref.~\cite{szankowski_accuracy_2018}, such that
\begin{equation}
    \chi=T\sum_m|c_m|^2\int_0^T dt (1-\frac{t}{T})G(|t|)e^{im\omega_p t}+\sum_{m_1\neq -m_2}\frac{c_{m_1}c_{m_2}}{(m_1+m_2)\omega_p}\int_0^T G(|t|)(e^{im_1\omega_p t}-e^{-im_2\omega_p t})
    \label{Supp_eq:chi_t}
\end{equation}
where $c_m=\frac1T \int_o^Te^{-im\omega_p t}f(t)dt$ is the Fourier component of filter function at frequency $m\omega_p$, where $\omega_p=\frac{2\pi}{T}$ and $f(t)=\sum_{m}c_m e^{im\omega_p t}$ ($0\leq t\leq T$). For large $T$ in comparison to the noise correlation time $\tau_c$, the ``off-diagonal" term (the second term) and $t/T$ term in the ``diagonal" term (the first term) in Eq.~\eqref{Supp_eq:chi_t} are neglected.

\subsection{Noise spectroscopy with CPMG sequences}

For CPMG sequences comprising $N$ $\pi$ pulses and pulse interval $T/N$ (note that $N$ is an even number), the filter function is calculated as
\begin{equation}
    |F(\omega)|^2=\frac{16\sin^2(\frac{N\omega\tau}{2})\sin^4(\frac{\omega\tau}{4})}{\omega^2\sin^2(\frac{\omega\tau}{2}-\frac\pi2)}
\end{equation}
The poles of $|F(\omega)|^2$ are $\omega=(2k+1)\pi/\tau$, where $k$ is any integer. In the limit of large $N$, we can approximately use Dirac $\delta$ functions to express $|F(\omega)|^2$, 
 \begin{align}
    |F(\omega)|^2\approx \sum_{k=-\infty}^\infty \frac{16}{\omega^2}\frac{\sin\left(\frac{N}{2}\,\omega\tau\right)^2}{(\omega\tau-(2k+1)\pi)^2} \approx \sum_{k=-\infty}^\infty \frac{8\pi N}{\omega^2}\delta\left(\omega\tau-(2k+1)\pi\right) ,
 \end{align}
 where we use the $\delta$-function approximation 
\begin{equation}
    \frac1{t}\frac{\sin^2(t(x-a))}{(x-a)^2}|_{t\rightarrow \infty}=\pi\delta(x-a),
\end{equation}
 and calculate the attenuation function $\chi$ as
\begin{align}
    \chi&=\frac12\int_{-\infty}^{+\infty}\frac{d\omega}{2\pi}S(\omega)|F(\omega)|^2\\
    &\approx  \frac12 \int_{-\infty}^{+\infty}\frac{d\omega}{2\pi}S(\omega)\frac{8\pi N}{\omega^2}\sum_{k=-\infty}^{+\infty}\delta\left(\omega\tau-(2k+1)\pi\right)\\
    &= \frac{2T}{\tau^2}\sum_{k=-\infty}^{+\infty} \int_{-\infty}^{+\infty}{d\omega}\frac{S(\omega)}{\omega^2}\delta\left(\omega-(2k+1)\pi/\tau\right)\\
    &=T\frac{4}{\pi^2}\sum_{k=0}^{+\infty}\frac{1}{(2k+1)^2}S(\frac{(2k+1)\pi}{\tau})\label{Supp_eq:CPMG_chi}
\end{align}
where we use the symmetry of the spectrum for classical noise, 
$S(\omega)=S(-\omega)$.

As derived, the accuracy of the noise spectroscopy relies on the approximation to Dirac comb structure of the filter function, which corresponds to a large $T$ or large $N$ limit. The expression in Eq.~\eqref{Supp_eq:chi_t} further elucidates such a criterion 
in the time domain. For large $T$ in comparison to the noise correlation time $\tau_c$, the ``off-diagonal" term and $t/T$ term in the ``diagonal" term in Eq.~\eqref{Supp_eq:chi_t} are neglected. For CPMG sequence with large $T$, we have 
\begin{equation}
    \chi=T\sum_{k=0}^{\infty} |c_k|^2S(\frac{(2k+1)\pi}{\tau}),
    \label{CPMGtrans}
\end{equation}
with
\begin{equation}
    |c_k|^2=\frac{16\sin^4(\frac{m\omega_p \tau}{4})\sin^2(\frac{\omega_p N\tau}{2})}{m^2\omega_p^2T^2 \cos^2(\frac{m\omega_p\tau}{2})}|_{m=2k+1}=\frac{4\sin^2(kN\pi)}{(2k+1)^2\pi^2N^2\sin^2({k\pi})}=\frac{4}{\pi^2}\frac{1}{(2k+1)^2},
\end{equation}
where $\omega_p=\pi/\tau$ and $m=(2k+1)$ ($k=0,1,2,\cdots$) are odd numbers due to the symmetry of CPMG sequences. In this way, we obtain the same $\chi$ as in Eq.~\eqref{Supp_eq:CPMG_chi}.

Similarly, for a Ramsey sequence, we can calculate the filter function
\begin{equation}
    |F(\omega)|^2=\frac{4\sin^2(\frac{\omega T}{2})}{\omega^2},
\end{equation}
and the attenuation exponent in the large $T$ limit reads
\begin{align}
    \chi&=\frac1\pi \int_{-\infty}^{+\infty}\frac{\sin^2(\frac{\omega T}{2})}{\omega^2}S(\omega)d\omega\\
    &\approx \frac1\pi \int_{-\infty}^{+\infty} \frac{T\pi}{2}\delta(\omega)S(\omega)d\omega\\
    &=\frac{T}{2}S(0).\label{Supp_eq:Ramsey_chi}
\end{align}

Both Eqs.~\eqref{Supp_eq:CPMG_chi} and \eqref{Supp_eq:Ramsey_chi} form the basis of standard comb-based noise spectroscopy. The noise spectrum components $S(0)$ and $S(\pi/\tau)$ can be directly reconstructed by measuring the attenuation function $\chi$.

\subsection{Noise spectroscopy with Walsh sequences}

Rather than measuring the noise spectrum by sweeping the frequency filter of CPMG sequences, our approach employes a ``direct" reconstruction of the correlation in the time domain with Walsh modulation, namely, a digital modulation scheme that affords intrinsic compatibility with 
hardware constraints
and time discretization~\cite{hayes_reducing_2011}. 

By applying Walsh modulation sequences, the attenuation function $\chi$ is mapped to the logical auto-correlation of the noise, which is derived based on the logical Wiener-Khintchine theorem~\cite{robinson_logical_1972}. The arithmetic auto-correlation is then obtained through a linear transformation, before we apply a Fourier transform to obtain the Fourier power spectrum of the noise. 

We denote the modulation function $f(t)$ of Walsh sequence $m$ as $f(t)=W_m(\lfloor t/\tau\rfloor)$ where $\tau=T/N$, and choose the ``sequency" labeling convention such that the pulse number (denoting value switch of $f(t)$) for the $m$-th sequence is $m$. The attenuation $\chi$ under the $m$-th walsh sequence is then calculated as
\begin{eqnarray}
    \chi_m & = &\frac12\int_0^T\int_0^T dt_1dt_2G(t_1-t_2)W_m(\lfloor \frac{t_1}{\tau}\rfloor )W_m(\lfloor \frac{t_2}{\tau}\rfloor )=\frac{T^2}{2N^2}\sum_{k,j}\Bar{G}_N(j-k)W_m(j)W_m(k) \\
    & \approx &\frac{T^2}{2N^2}\sum_{k,j}{G}_N(j-k)W_m(j)W_m(k),
\end{eqnarray}
where the continuous time from $0$ to $T$ is discretized to $N$ equal pieces and the arithmetic auto-correlation is approximated with
\begin{equation}
    \Bar{G}_N(j-k)=\frac{N^2}{T^2}\int_{kT/N}^{(k+1)T/N}\int_{jT/N}^{(j+1)T/N}dt_1dt_2G(t_1-t_2)\approx G_N(j-k).
\end{equation}

We also define the local logical auto-correlation for the discretized noise $\omega(j)$ as~\cite{robinson_logical_1972}
\begin{equation}
    L_N(j)=\frac1N \sum_{k=0}^{N-1} \langle \omega(k)\omega(k\oplus j)\rangle=\frac1N \sum_{k=0}^{N-1} G_N(j\oplus k -k),
\end{equation}
where $\oplus$ denotes the bitwise modulo 2 addition of the two integers $j$ and $k$.
The reason why we use bit-by-bit binary modulo is that the walsh sequences are defined under the binary basis, which satisfies the dyadic composition equality $W_m(j)W_m(k)=W_m(j\oplus k)$. Thus we have 
\begin{equation}
    \chi_m\approx\frac{T^2}{2N^2}\sum_{j,k} G_N(j-k) W_m(j\oplus k)=\frac{T^2}{2N^2}\sum_{j,k} G_N(j\oplus k-k) W_m(j)=\frac{T^2}{2N}\sum_j L_N(j)W_m(j),
\end{equation}
which directly connects the logical auto-correlation with the attenuation function through a linear transformation $W^{-1}$ (the inverse of the Walsh matrix). The arithmetic auto-correlation is then obtained from the logical auto-correlation with a linear transformation~\cite{robinson_logical_1972}
\begin{equation}
    L_N(k)=D_N(k,k)\sum_jT_N(k,j)G_N(j), \quad G_N(j)=\sum_kT_N^{-1}(k,j)D_N^{-1}(k,k)L_N(k).
\end{equation}

\subsubsection*{Walsh reconstruction protocol}

In summary, the protocol for the reconstruction of the noise spectrum is as follows:
\begin{enumerate}
    \item Measure $\chi_m$ for $m=0,1,2,\cdots,N-1$;
    \item Calculate the discrete logical auto-correlation $L_N(j)$ for $j=0,1,2,\cdots,N-1$ with
    \begin{equation}
        L_N(j)=\frac{2N}{T^2}\sum_m W_N^{-1}(j,m)\chi_m;
        \label{walshchi2L}
    \end{equation}
    \item Calculate the discrete arithmetic auto-correlation $G_N(j)$ with
    \begin{equation}
        G_N(j)=\sum_kT_N^{-1}(k,j)D_N^{-1}(k,k)L_N(k);\
        \label{walshL2G}
    \end{equation}
    \item Calculate the noise spectrum $S(\omega)$ with discrete Fourier transform of the arithmetic auto-correlation.
\end{enumerate}

As derived, the accuracy of the auto-correlation reconstruction depends only 
upon the discretization of the time from $0$ to $T$. For standard noise spectroscopy, the accuracy is also affected by the discretization (and truncation) of the Fourier transform. As discussed in the main text and Appendix B, the above digital reconstruction protocol may also be reinterpreted within the general formalism of frame-based noise characterization \cite{Chalermpusitarak21}.

\subsubsection*{Generation of $T_N$, $D_N$ matrices}

Although Ref.~\cite{robinson_logical_1972} reports details on generating the transformation matrices such as $T_N$ and $D_N$, here we briefly summarize the derivation,  in order to make the presentation more self-contained. 

The bi-linear transfer matrix $T_N$ can be recursively generated as follows:
\begin{align}
    T_N &=
    \begin{pmatrix}
    T_{N/2} & 0 \\
    T_{N/2}S_{N/2} &  T_{N/2} 
    \end{pmatrix},
    \label{Supp_eq:TN}
\end{align}
where $S_N$ is the $N\times N$ shuffling matrix with unit elements ``off to the right of the SW-NE diagonal" such as $S_1=0$, $S_2=
\begin{pmatrix}
    0 & 0 \\
    0 & 1
\end{pmatrix}$, and $S_4=
    \begin{pmatrix}
    0 & 0 & 0 & 0\\
    0 & 0 & 0 & 1\\
    0 & 0 & 1 & 0\\
    0 & 1 & 0 & 0    
    \end{pmatrix}$. 
Based on the initial values $T_1=1$, we obtain 
\begin{align}
    T_2 &=
    {
    \begin{pmatrix}
    T_{1} & 0 \\
    T_{1}S_{1} &  T_{1} 
    \end{pmatrix}
    }=
    {
    \begin{pmatrix}
    1 & 0 \\
    0 & 1 
    \end{pmatrix}.
    }
    \label{Supp_eq:T2}
\end{align}
Based on $T_2$ we get above, we can get $T_4$ as:
\begin{align}
    T_4 &=
    {
    \begin{pmatrix}
    T_{2} & 0 \\
    T_{2}S_{2} &  T_{2} 
    \end{pmatrix}
    }=
    {
    \begin{pmatrix}
    1 & 0 & 0 & 0\\
    0 & 1 & 0 & 0\\
    0 & 0 & 1 & 0\\
    0 & 1 & 0 & 1
    \end{pmatrix}.
    }
    \label{Supp_eq:T4}
\end{align}  
Then the transfer matrices $T_8$, $T_{16}$, $T_{32}$, etc. can be obtained recursively.

In turn, the diagonal matrix $D_N$ can be generated with $D_N(k,k)=2^{1-\delta(k,0)-V_k}$, where $\delta(k,0)=1$ if and only if k=0, $V_k$ is the number of ones of $k$ represented in the binary format.

\section{Comparison between Walsh and CPMG noise spectroscopy}
\subsection{Discrete Fourier transform and comparison scheme}

\begin{figure}[h]
\centering \includegraphics[width=0.8\textwidth]{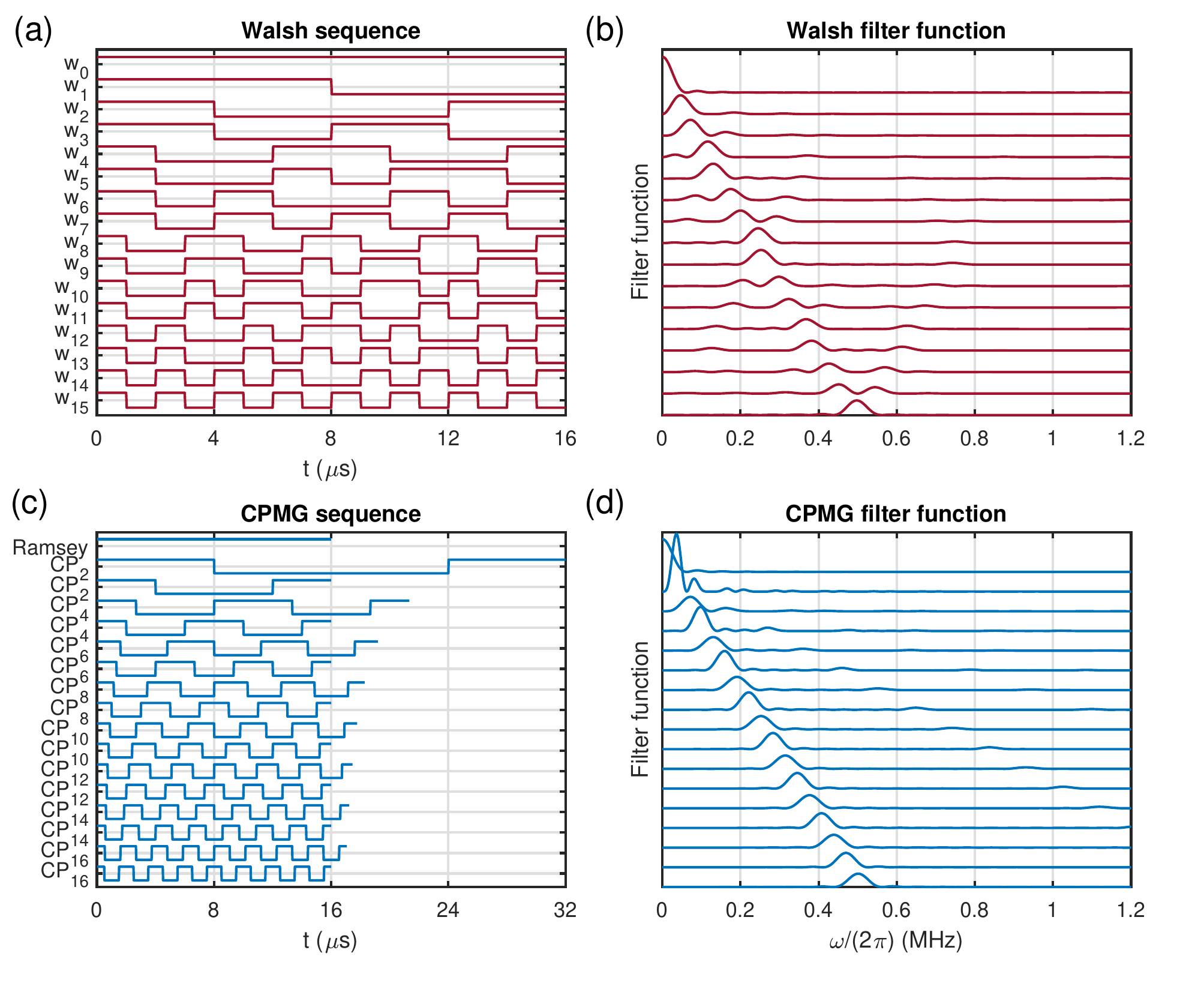}
\caption{\label{Supp_Fig_WalshCPMG} An example of the comparison scheme between Walsh and CPMG sequences with  $N=32$ and $T=16$ $\mu s$.}
\end{figure}

Since the reconstructed Walsh auto-correlation or CPMG spectrum are values of $G(t)$ or $S(\omega)$ at specific time or frequency points (rather than average values in some time or frequency intervals), in this work we modify a bit the discrete Fourier transform algorithm to make it compatible with our protocols and to give less errors than existing fast Fourier transform functions (such as ``fft" functions in MATLAB).

For the noise reconstruction with the Walsh method using $N=2^n$ sequences with time $T$, we first obtain discrete values of the auto-correlation $G(t_j)$ with $t_j=0,\frac TN,\cdots,\frac{T(N-1)}{N}$. With the assumption of a symmetric auto-correlation such that $G(-t)=G(t)$, we can actually extend the time range and obtain $2N-1$ discrete $G(t_j)$ points with $t_j=jT/N$, $j=-(N-1),\cdots,(N-1)$. The corresponding discrete Fourier frequency components we can obtain is then $S(\omega_k)$, with $\omega_k=\frac{\pi k}{T(N-1)/N}$ ($k=0,\cdots,N-1$), which can be calculated using
\begin{align}
    S(\omega_k)=\int_{-\infty}^{\infty}G(t)e^{-i\omega_k t}dt\approx \sum_{j=-(N-1)}^{N-1}G(t_j) e^{-i\pi\frac{k}{T(N-1)/N} \frac{jT}{N}}\Delta t=\sum_{j=-(N-1)}^{N-1}G(t_j) e^{-i\pi\frac{kj}{N-1}}\Delta t ,
\end{align}
where $\Delta t=T/N$.

To make a fair comparison, the first and last sequences for the CPMG method sample the same frequency as the Walsh method such that the ranges of reconstructed spectrum and auto-correlation are the same as (or similar to) the Walsh method. In addition, the total sequence time $\sim NT$ for both methods should also be similar to each other. More precisely, the first sequence of CPMG scheme is a Ramsey sequence with $f(t)=1$, and the $2k,2k+1$ sequences are both CPMG-$k$ sequences with pulse interval $\tau_{2k}=T/(2k-1)$, $\tau_{2k+1}=T/(2k)$, respectively. Note that $k$ takes values from $1$ to $N/2$ (see an example of the comparison shown in Fig.~\ref{Supp_Fig_WalshCPMG}). Such a set of CPMG sequences give a sampling in frequency space $\omega_k=k\pi/T$ with $k=0,\cdots,N$. As long as 
$S(-\omega)=S(\omega)$, we can extend the frequency range and obtain $S(\omega_k)$ with $2N+1$ points $\omega_k=-\frac{\pi N}{T},\cdots,-\frac{\pi}{T},0,\frac{\pi}{T},\cdots,\frac{\pi N}{T}$. The corresponding discrete auto-correlation we could obtain is then $G(t_j)$ with $t_j=\frac{jT}{N}$ ($j=0,\cdots,N$), which can be calculated using
\begin{align}
    G(t_j)=\frac{1}{2\pi}\int_{-\infty}^{\infty}S(\omega)e^{i\omega t_j}d\omega &\approx\frac{1}{2\pi}\sum_{k=-N}^{N} S(\omega_k) e^{i\pi\frac{k}{T} \frac{jT}{N}}\Delta\omega=\sum_{k=-N}^{N} S(\omega_k) e^{i\pi\frac{kj}{N}}\frac{\Delta\omega}{2\pi}.
\end{align}

\subsection{Ornstein-Uhlenbeck noise}

The characteristic spectrum of a spin qubit dipolarly coupled to spin bath can be described by the Ornstein-Unlenbeck (OU) process~\cite{de_lange_universal_2010}. The time-dependent noise field $\omega(t)$ of the OU noise has an auto-correlation function of the form
\begin{equation}
    G(t)=\langle\omega(t)\omega(0)\rangle=b^2e^{-\frac{t}{\tau_c}}
\end{equation}
where $b^2$ and $\tau_c$ characterize the noise strength and the correlation time, respectively. The corresponding noise spectrum is then 
\begin{equation}
    S(\omega)=\int_{-\infty}^{\infty}dtb^2e^{-\frac{|t|}{\tau_c}}e^{-i\omega t}=\frac{2b^2\tau_c}{1+\omega^2\tau_c^2}, 
\end{equation}
that is, a Lorentzian lineshape.

We further consider the noise source with a nonzero frequency $\omega_s$, such that the auto-correlation is modified to 
\begin{equation}
    G(t)=\langle\omega(t)\omega(0)\rangle=b^2e^{-\frac{t}{\tau_c}}\cos(\omega_st)
\end{equation}
and the corresponding spectrum is 
\begin{equation}
    S(\omega)=\frac{b^2\tau_c}{1+(\omega-\omega_s)^2\tau_c^2}+\frac{b^2\tau_c}{1+(\omega+\omega_s)^2\tau_c^2}.
    \label{Supp_eq:S_OU_theory}
\end{equation}


\subsection{Metric to evaluate the noise reconstruction}

Walsh sequence reconstructs the auto-correlation of the noise directly using the correlation of the attenuation functions under different sequences, and the noise spectrum is then obtained through a discrete Fourier transform. While for CPMG sequence, the noise spectrum is first reconstructed before transforming to the auto-correlation in time space. Since the discrete Fourier transform gives rise to error due to its imperfect approximation to Fourier transform, we will need to evaluate the reconstruction both in time and frequency space, and make them comparable through proper normalization. Here we define two metrics to quantify the average error and individual error in the noise reconstruction.

For the reconstructed noise parameter $A$ and its theoretical value $A_0$ (here $A$ can be noise spectrum $S$ and noise auto-correlation $G$), we define the average error as \begin{equation}
    \epsilon(A)=\frac{\sum_{i} (A(i)-A_0(i))^2}{\sum_i A_0(i)^2}.
\end{equation}
Sometimes we are interested in discussing the error of noise reconstruction in different frequency or time regions, thus we also define the error for an individual reconstructed point as
\begin{equation}
    E(A(i))=\bigg|\frac{A(i)-A_0(i)}{A_0(i)}\bigg|, 
\end{equation}
which is similar 
to the definition of $\chi$ error in Ref.~\cite{szankowski_accuracy_2018} Eq.~(20).

\subsection{Comparison}

As discussed above, for a comparison between Walsh and CPMG with fixed sequence time $T$, the sampling of both methods in time and frequency domain are similar. For an OU noise at frequency $\omega_s$ with correlation time $\tau_c$, we find the following results for the average reconstruction error:

\begin{enumerate}
    \item Walsh spectroscopy: 
    \begin{itemize}
    \item $G(t)$ is first reconstructed almost perfectly and the error is dominated by the relation between the time space sampling $T/N$ and the characteristic noise properties (the ``average" curvature of the auto-correlation). 
    
    For an OU noise at frequency $\omega_s$, the characteristic timescale is given by correlation time $\tau_c$ and the noise period $T_s=2\pi/\omega_s$.
    
    \item $S(\omega)$ is reconstructed through discrete Fourier transform of $G(t)$. In addition to linearly propagated error of $G(t)$, the discrete Fourier transform introduces error dominated by the relation between $\tau_c$ and $T$, as well as the relation between $\tau_c$ and $T_s$.
    \end{itemize} 
        
    \item CPMG spectroscopy:
    \begin{itemize}
    \item $S(\omega)$ is first reconstructed and the error is dominated by the relation between the frequency space sampling $\pi/T$ and the characteristic spectrum properties including noise frequencies and linewidths. A significant error source comes from the $\delta$-function approximation of the filter functions, which is eliminated when $\tau_c/T\ll 1$.
    \item $G(t)$ is reconstructed through discrete Fourier transform of $S(\omega)$. In addition to linearly propagated error of $S(\omega)$, the discrete Fourier transform introduces error dominated by the relation between $\pi/T$ and the characteristic spectrum properties including noise frequencies $\omega_s$, linewidths $1/\tau_c$, as well as the frequency sampling range.   
    \end{itemize}
\end{enumerate}

Besides the overall performance of the noise reconstruction discussed in the main text, here we show an example where we look at the error of individual points in the reconstructed spectrum or auto-correlation, as defined above. 
Specifically, in Fig.~\ref{Supp_Fig_IndividualError_tau1} we show an example of the spectrum reconstruction error, where the Walsh method shows larger error at larger frequency due to the insufficient time space sampling. When $n$ increases, the error for Walsh decreases for all frequencies, while the error for CPMG decreases only for larger frequency due to the effect of taking into account more higher harmonics for the high-frequency components.

\begin{figure}[htbp]
\centering \includegraphics[width=0.5\textwidth]{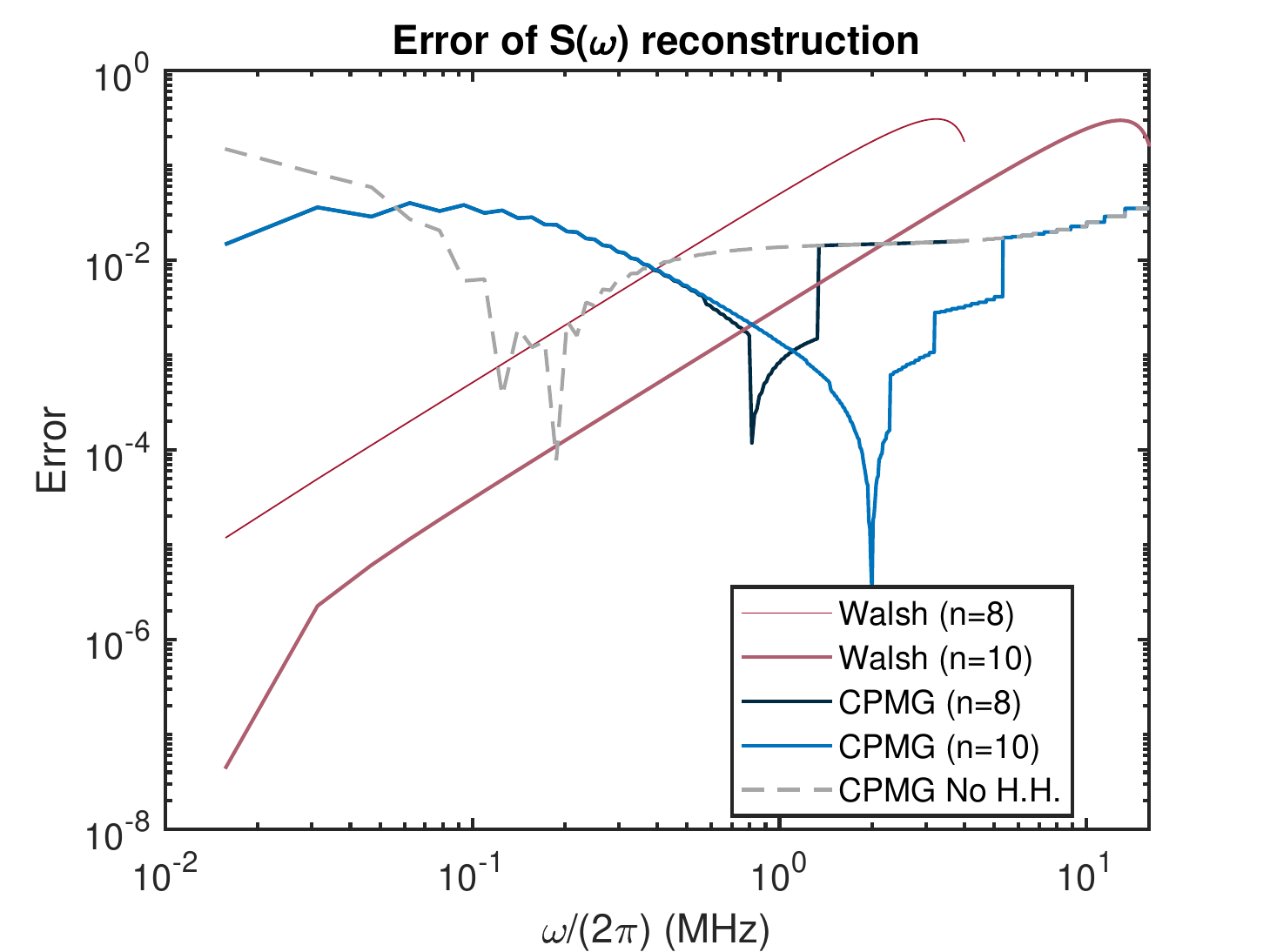}
\caption{\label{Supp_Fig_IndividualError_tau1} Error of individual points in the reconstructed noise spectrum using CPMG and Walsh methods. The OU noise parameters used are $\tau=1\ \mu\text{s},\omega_s=0,b^2=0.003125\ \mu$s.  }
\end{figure}

\section{Statistical analysis of the OU noise model}

We note that in the reconstruction comparison discussed in the main text, we simulate the attenuation functions by directly calculating the integral in Eq.~\eqref{Supp_eq:chi_theory}. However, those simulations are based on an assumption of infinite number of experimental averages or noise instances, which need not be satisfied in practical applications and does not give enough insight about the possible errors induced by the stochastic nature of the noise. Thus, we add a statistical analysis of the OU noise model and analyze the potential impact of insufficient experimental averages or noise instances.

To generate a static OU noise that satisfies a stationary Gaussian distribution, we can use an ``updating" equation~\cite{gillespie_exact_1996,gillespie_mathematics_1996}, namely, 
\begin{equation}
    \omega(t+\delta t)=\omega(t)e^{-\frac{dt}{\tau_c}}+b\sqrt{1-e^{-\frac{2dt}{\tau_c}}}r_1,
\end{equation}
where $r_1$ is a random number satisfying a Gaussian distribution $r_1\sim N(0,1)$. By updating the time trace of the noise in the same fashion, we obtain
\begin{align}
    \omega(t+2\delta t)&=\omega(t+dt)e^{-\frac{dt}{\tau_c}}+b\sqrt{1-e^{-\frac{2dt}{\tau_c}}}r_1=\omega(t)e^{-2\frac{dt}{\tau_c}}+b\sqrt{1-e^{-\frac{2dt}{\tau_c}}}r_2e^{-\frac{dt}{\tau_c}}+b\sqrt{1-e^{-\frac{2dt}{\tau_c}}}r_1 , \\\cdots\\
    \omega(t+k\delta t) &=  \omega(t)e^{-k\frac{dt}{\tau_c}}+b\sqrt{1-e^{-\frac{2dt}{\tau_c}}}\left(r_1+r_2e^{-\frac{dt}{\tau_c}}+\cdots+r_ke^{-(k-1)\frac{dt}{\tau_c}}\right).
\end{align}
Thus, the auto-correlation $G(kdt)$ is
\begin{equation}
    G(kdt)=\langle\omega(t+kdt)\omega(t)\rangle=b^2e^{-\frac{kdt}{\tau_c}},
    \label{Supp_eq:G(kdt)}
\end{equation}
where we use the fact that the square of a normal distribution $\omega(t)/b\sim {\cal N}(0,1)$ is effectively a chi-squared distribution, $\omega(t)^2/b^2\sim \chi_1^2$.
Note that when the noise trace is generated from an initial time $t_0$ with a constant value $\omega(t_0)=\text{const.}$, the correlation between times $t$ and $t+kdt$ is $\langle\omega(t+kdt)\omega(t)\rangle=b^2e^{-\frac{kdt}{\tau_c}}(1-e^{-2\frac{t-t_0}{\tau_c}})$, which is reduced to Eq.~\eqref{Supp_eq:G(kdt)} only when $t-t_0\gg\tau_c$~\cite{gillespie_mathematics_1996}. In practical simulation, we can set the initial point with a random number $\omega(0)=br_1$ satisfying a Gaussian distribution to avoid such a problem.

We note that Eq.~\eqref{Supp_eq:G(kdt)} is obtained under the assumption that the average is taken on a sufficiently large number of noise realizations. For practical applications, it is crucial to lower bound the number of averages needed. We analyze Eq.~\eqref{Supp_eq:G(kdt)} again for a single measurement (or noise instance), and the ``auto-correlation" is actually
\begin{align}
    \omega(t+kdt)\omega(t)&=\omega(t)^2e^{-\frac{kdt}{\tau_c}}+b\sqrt{1-e^{-\frac{2dt}{\tau_c}}}\omega(t)(r_1+r_2e^{-\frac{dt}{\tau_c}}+\cdots+r_k e^{-(k-1)\frac{dt}{\tau_c}})\label{Supp_eq:autocorr}
    \\
    &=b^2e^{-\frac{kdt}{\tau_c}}+\epsilon_0+\epsilon_1 ,
\end{align}
where $\langle \omega(t+kdt)\omega(t)\rangle=b^2e^{-kdt/\tau_c}$, $\epsilon_0$ and $\epsilon_1$ are zero-mean random variables corresponding to the first and second terms in Eq.~\eqref{Supp_eq:autocorr}. The variance of the summation $\epsilon_0+\epsilon_1$ can be calculated using the following steps:
\begin{align}
    var(\epsilon_0)&=2b^4e^{-\frac{2kdt}{\tau_c}},\\ var(\epsilon_1)&=b^4({1-e^{-\frac{2dt}{\tau_c}}})(1+e^{-\frac{2dt}{\tau_c}}+e^{-\frac{4dt}{\tau_c}}+\cdots+e^{-\frac{2(k-1)dt}{\tau_c}})\\&=b^4 ({1-e^{-\frac{2dt}{\tau_c}}})\frac{1-e^{-\frac{2kdt}{\tau_c}}}{1-e^{-\frac{2dt}{\tau_c}}}\\&=b^4 ({1-e^{-\frac{2kdt}{\tau_c}}}),\\
    cov(\epsilon_0,\epsilon_1)&=0,
    \\var(\epsilon_0+\epsilon_1)&=var(\epsilon_0)+var(\epsilon_1)+cov(\epsilon_0,\epsilon_1)=b^4(1+e^{-\frac{2kdt}{\tau_c}}).
\end{align}

Thus, when we consider a finite number of averages, denoted by $\langle \:\rangle_N$ ($N$ averages), the variance of the auto-correlation is 
\begin{equation}
    var(\langle \omega(t_1)\omega(t_2)\rangle_N)=\frac{b^4(1+e^{-\frac{2|t_1-t_2|}{\tau_c}})}{N}.
    \label{Supp_eq:variance_N}
\end{equation}
Accordingly, the ratio of the standard deviation to the auto-correlation is
\begin{equation}
    \frac{std(\langle \omega(t_1)\omega(t_2)\rangle_N)}{\langle \omega(t_1)\omega(t_2)\rangle}=\sqrt{\frac{1+e^{\frac{2|t_1-t_2|}{\tau_c}}}{{N}}},
    \label{Supp_eq:std_N}
\end{equation}
which increases with the time difference.
These relations are validated numerically in Fig.~\ref{Supp_Fig_VarianceGt}, where we simulate an OU noise and plot both the simulated and the theoretically predicted variance of the auto-correlation $\langle\omega(t)\omega(0)\rangle_N$. For practical applications, Eqs.~\eqref{Supp_eq:variance_N} and~\eqref{Supp_eq:std_N} are useful for judging whether the number of averages is sufficient for estimation purposes.
In particular, when the number of averages is not sufficient, the relative error in the Walsh reconstruction may be instead dominated by the fluctuations of the noise source itself, especially for those points with larger $t$ in auto-correlation $G(t)$.

\begin{figure}[h]
\centering \includegraphics[width=0.7\textwidth]{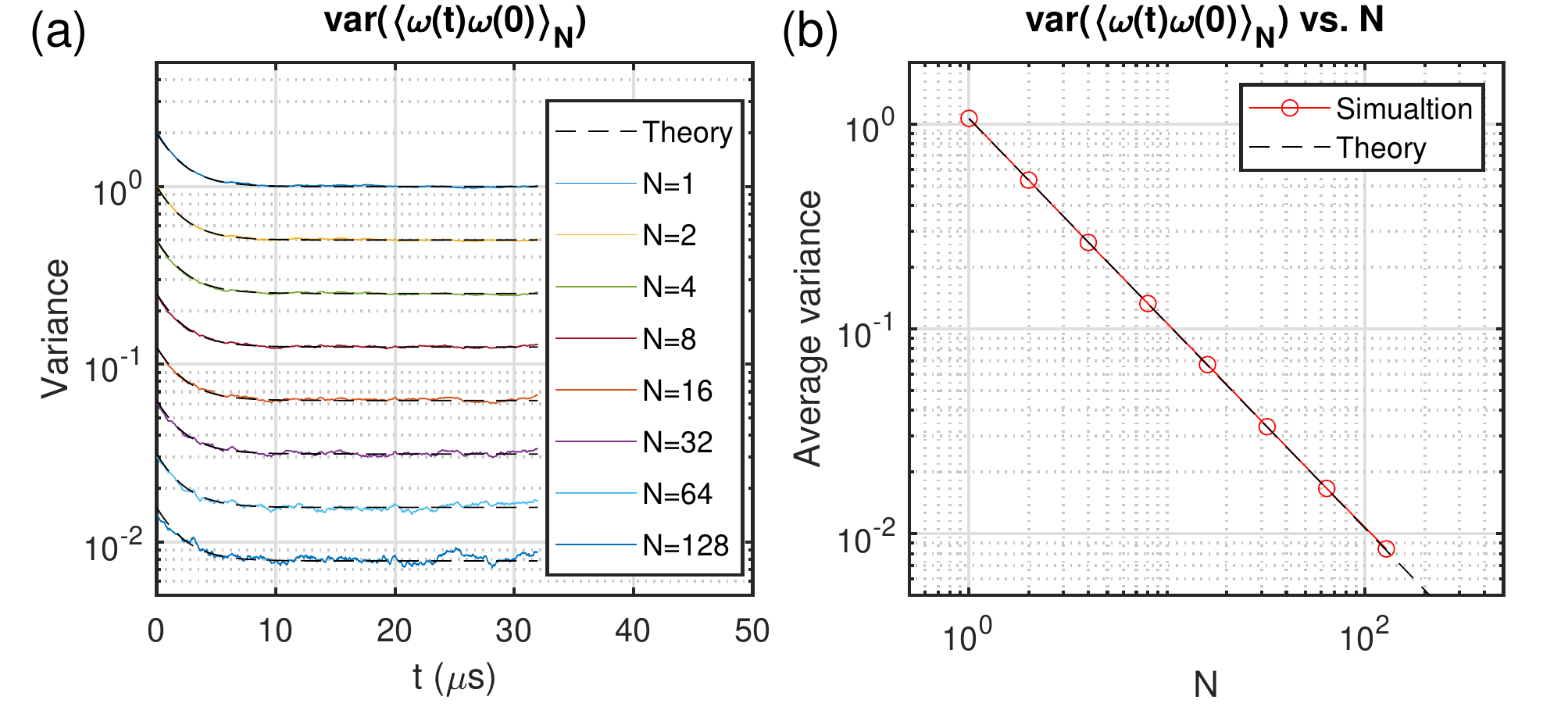}
\caption{\label{Supp_Fig_VarianceGt} Variance of $\omega(t)\omega(0)$ in a simulated OU noise. (a) Variance of $G(t)$ in a simulated static OU noise trace with $b^2=1\ \mu$s, $\tau_c=4\ \mu$s, $\omega_s=0$. We use a time step $0.03125\ \mu$s and simulate the noise from $0$ to $32\ \mu$s with $65536=2^{16}$ repetitions. The simulated noise traces are then divided into $2^{16-n}$ groups with size $N=2^n$ such that the variance of the mean value of different group size can be calculated. The theory curve is plotted using function $b^4(1+e^{-2t/\tau_c})/N$ (b) The mean value of variances (over all points in the time traces for each $N$) in (a) plotted as a function of $N$.}
\end{figure}

The analysis above only sets a theoretical upper bound, instead of a precise value, for the error in auto-correlation reconstruction with the Walsh method. The discretization of the continuous auto-correlation takes the average value of $\omega(t_j)\omega(t_k)$ for $t_j\in [jT/N,(j+1)T/N], t_k\in [kT/N,(k+1)T/N]$ as the $G_N(j-k)$, which potentially achieves an average effect and partly decreases the influence of insufficient average. 
In addition, under the stationary noise assumption, the auto-correlation is only dependent on the time difference, thus its reconstruction is an effective ``average" of many ``time pairs" in the time trace with the same time difference, which potentially yields less reconstruction errors as well. We note that although these effects could potentially allow less average for a satisfying reconstruction, in Walsh experiments the minimal inter-pulse delay is usually set to be smaller than the noise correlation time (for satisfying sampling in time domain), and the improvement of both aforementioned ``average" effects is minor due to the strong correlation within each inter-pulse delay.

\section{Quantum-classical environment}
\label{Supp_DerivationQCmodel}

Although our theoretical analysis and experimental demonstration are based on the assumption of a classical noise model satisfying Gaussian distribution, in a more general case the noise source of a qubit comprises both a few nearby quantum spins and a far-away classical bath (which might be itself of  quantum nature, but able to be reduced to a classical model in view of its large size). Then the task is in principle to both detect the quantum spins and the classical bath. The method to extract information about a quantum-classical environment has been well-developed using CPMG or XY sequences~\cite{kolkowitz_sensing_2012,hernandez-gomez_noise_2018,childress_coherent_2006,taminiau_detection_2012}. However, a precise detection of the coupling constant due to the quantum spin requires sweeping the number of pulses to obtain the population oscillation of an initial state (say,  $\ket{+}$). In contrast, both the Walsh and CPMG methods studied in this work, with a ``single" point measurement atfixed sequence time $T$, are designed to reconstruct a classical noise and are thus unable to separate contributions from the quantum and classical noise sources easily. As mentioned above, the ``single" point CPMG sequence can be easily modified to extract the environmental information (in particular, the quantum part), so here we briefly discuss and compare the error introduced by the quantum spin when reconstructing the classical part for both Walsh and CPMG methods. We note that a more in-depth study, while being of interest for future investigation,  is beyond the scope of this work.

\subsection{Derivation of the system dynamics under DD}

We consider the NV electronic spin coupled to a $^{13}$C nuclear spin bath. Under the secular approximation due to the large electronic energy scale, the Hamiltonian is written as
\begin{equation}
    H=\sum_j S_zA_{\parallel}^{(j)}I_z^{(j)}+S_zA_{\perp}^{(j)}I_x^{(j)}+\omega_L^{(j)}I_z^{(j)},
\end{equation}
where $j$ denotes the $j^{\text{th}}$ nuclear spin in the bath with a Larmor frequency $\omega_L^{(j)}$, and $A_{\parallel}^{(j)},A_\perp^{(j)}$ are the parallel and perpendicular hyperfine strengths, respectively. The magnetic field is aligned along the NV axis and the nuclear spin of $^{14}$N is neglected.

The evolution of each nuclear spin is dependent on the state of the electron spin. We use the two NV spin states $\ket{m_s=0,-1}$ (denoted as $\ket{0},\ket{1}$), such that the Hamiltonian can be further written as 
\begin{equation}
    H=\sum_j \ket{0}\bra{0}\otimes H_{0}^{(j)}+\ket{1}\bra{1}\otimes H_{1}^{(j)}, 
\end{equation}
where the NV-dependent nuclear spin operators are
\begin{equation}
    H_{0}^{(j)}=\omega_L^{(j)}I_z^{(j)}; \quad H_{1}^{(j)}=(\omega_L-A_{\parallel}^{(j)})I_z^{(j)}-A_{\perp}^{(j)}I_x^{(j)}.
\end{equation}
Note that here we do not consider the intra-bath interaction of different nuclear spins; accordingly, their dynamics are separable.

For most DD experiments, the initial state of the NV center is prepared to $\ket{+}$, and we assume the nuclear spins are fully mixed states, such that the joint initial density matrix is
\begin{equation}
    \rho_0=\ket{+}\bra{+}\otimes \frac12 \mathds{1}^{(1)}\otimes \frac12 \mathds{1}^{(2)}\otimes\cdots.
\end{equation}
The state of the system at time $T$ is then 
\begin{equation}
    \rho(t)=U(T)\rho_0U^\dagger(T),
\end{equation}
where the unitary evolution operator is $U(T)=\mathcal{T}\exp(-i\int_0^T H(t)dt).$ To deal with such an operator, one can either expand it through a perturbative Magnus expansion~\cite{hernandez-gomez_noise_2018}, or continue to analyze the spin-dependent evolution~\cite{taminiau_detection_2012}. Here we follow the latter method and discuss the system evolution under a DD sequence applied to the NV electronic spin, in such a way that the time from $0$ to $T$ is separated into $N+1$ pieces with duration $\tau_1,\tau_2,\tau_3,\cdots,\tau_{N+1}$ by $N$ instantaneous spin-flip pulses. Then $U(T)$ can be expressed as
\begin{equation}
    U(T)= \begin{cases}
\ket{0}\bra{0}\otimes U_0+\ket{1}\bra{1}\otimes U_1, & \text{$N$ even}, \\
\ket{1}\bra{0}\otimes U_0+\ket{0}\bra{1}\otimes U_1, & \text{$N$ odd}, \\
\end{cases}
\end{equation}
where the conditional evolution operators acting on the nuclear spins are 
\begin{align}
    U_0&=U_0^{(1)}\otimes U_0^{(2)}\otimes\cdots, \text{where } U_0{(j)}=\left[\exp(-iH_{\frac{1+(-1)^{N+1}}{2}}^{(j)}\tau_{N+1})\cdots \exp(-iH_{1}^{(j)}\tau_2)\exp(-iH_{0}^{(j)}\tau_1)\right],\\
    U_1&=U_1^{(1)}\otimes U_1^{(2)}\otimes\cdots, \text{where } U_1{(j)}=\left[\exp(-iH_{\frac{1-(-1)^{N+1}}{2}}^{(j)}\tau_{N+1})\cdots \exp(-iH_{0}^{(j)}\tau_2)\exp(-iH_{1}^{(j)}\tau_1)\right].
\end{align}
The signal is read out by measuring the NV population in $\ket{+}$ at time $T$, such that 
\begin{align}
    {\cal S}(T)=P_{\ket{+}}(T)=\frac12 \bigg[1+\Tr_{n}[\rho(T)(\ket{0}\bra{1}+\ket{1}\bra{0})]\bigg]=\frac12\left[1+\Re\left( \prod_j \frac12\Tr_{n}^{(j)}[U_0^{(j)}(T)U_1^{(j)\dagger} (T)]\right)\right].
\end{align}
Now the task is simply to calculate the signal contribution of each coupled nuclear spin
\begin{equation}
    M^{(j)}(T)=\frac12\Tr_{n}^{(j)}[U_0^{(j)}(T)U_1^{(j)\dagger} (T)],
\end{equation}
and then to obtain the signal
\begin{equation}
    {\cal S}(T)=\frac12\left[1+\Re\left(\prod_j M^{(j)}(T)\right)\right].
\end{equation}

If we include a classical noise bath to the model, the above signal is then modified to 
\begin{equation}
    {\cal S}(T)=\frac12\left[1+e^{-\chi(T)}\Re\left(\prod_j M^{(j)}(T)\right)\right],
    \label{Supp_eq:Signal}
\end{equation}
where $\chi(T)$ is the attenuation function we discussed in this work. Now in Eq.~\eqref{Supp_eq:Signal} we have obtained the signal of the NV electronic spin coupled to both quantum and classical noise sources.

\subsubsection*{Derivation for CPMG and XY sequences}

When a CPMG, XY, or periodic DD sequence is applied, the formula for $M^{(j)}(T)$ can be analytically calculated. Here we briefly summarize the derivation for CPMG and XY sequences. Note that, for simplicity, we hereby eliminate the subscript $(j)$. 

Following the derivation of Ref.~\cite{kolkowitz_sensing_2012}, we first calculate the evolution operator $V_0,V_1$ of a period with a $\tau/2-\tau-\tau/2$ sequence (CPMG-2), and then the complete evolution operators are $U_{0,1}=(V_{0,1})^{\frac{N}{2}}$. We define two angles $\phi_0=\frac{\omega_L\tau}{2}$, $\phi_1=\frac{\Tilde{\omega}\tau}{2}$ with $\Tilde{\omega}=\sqrt{(\omega_L-A_\parallel)^2+A_\perp^{(j)2}}$ and rotation axis $\mathbf{m_0}=\hat{z}$, $\mathbf{m_1}=\frac{\omega_L-A_\parallel}{\Tilde{\omega}}\hat{z}-\frac{A_\perp}{\Tilde{\omega}}\hat{x}$, and obtain the evolution operators 
\begin{align}
    V_{0,1}&=\exp[-i\frac{\phi_{0,1}}{2}\mathbf{m_{0,1}}\cdot\mathbf{\sigma}]\exp\bigg[-i\phi_{1,0}\mathbf{m_{1,0}}\cdot\mathbf{\sigma}\bigg]\exp[-i\frac{\phi_{0,1}}{2}\mathbf{m_{0,1}}\cdot\mathbf{\sigma}]
    =\cos(\varphi)\mathds{1}+i\sigma\cdot\mathbf{n_{0,1}}\sin(\varphi),
\end{align}
where $\mathbf{n_0},\mathbf{n_1}$, $\varphi$ are the rotation axes and angle for the two operators, given by
\begin{align}
    \cos\varphi&=\cos\phi_0\cos\phi_1-\mathbf{m_0}\cdot\mathbf{m_1}\sin\phi_0\sin\phi_1,\\
    \mathbf{n_0}&=\frac{\mathbf{m_0}(-\sin\phi_0\cos\phi_1+\sin\phi_1(1-\cos\phi_0)\mathbf{m_0}\cdot\mathbf{m_1})-\mathbf{m_1}\sin\phi_1}{\sin\varphi},\\
    \mathbf{n_1}&=\frac{\mathbf{m_1}(-\sin\phi_1\cos\phi_0+\sin\phi_0(1-\cos\phi_1)\mathbf{m_1}\cdot\mathbf{m_0})-\mathbf{m_0}\sin\phi_0}{\sin\varphi}.
\end{align}
The inner product between the two axes are then
\begin{equation}
    \mathbf{n_0}\cdot\mathbf{n_1}=1-\frac{A_\perp^2}{\Tilde{\omega}^2}\frac{(1-\cos\phi_0)(1-\cos\phi_1)}{1+\cos\phi_0\cos\phi_1-\frac{\omega_L-A_\parallel}{\Tilde{\omega}}\sin\phi_0\sin\phi_1}.
\end{equation}
Then we obtain the value of $M(T)$ as
\begin{align}
    M(T)=\frac12 \Tr(V_0^{\frac{N}{2}}V_1^{\frac{N}{2}\dagger})&=
    1-(1-\mathbf{n_0}\cdot\mathbf{n_1})\sin^2(\frac{N\phi}{2})=1-2\frac{A_\perp^2}{\Tilde{\omega}^2}\sin^2\left(\frac{\phi_0}{2}\right)\sin^2\left(\frac{\phi_1}{2}\right)\frac{\sin^2\left(\frac{N\varphi}{2}\right)}{\cos^2\left(\frac{\varphi}{2}\right)}.\label{Supp_eq:M(T)}
\end{align}

Typically, when the pulse spacing $\tau$ are fixed and the pulse number is swept, a coherent oscillation is observed due to term $\sin^2\frac{N\phi}{2}$ in Eq.~\eqref{Supp_eq:M(T)}, while the amplitude of the oscillation is highly sensitive to $\tau$ through parameters $\phi_0,\phi_1$ and $\varphi$. 
When $A\ll\omega_L$, such a resonance is obtained for $\phi_0\approx\phi_1=\pi$, 
yielding
an oscillation rate  proportional to the perpendicular component of the hyperfine constant,  
\begin{equation}
    M(T)=\cos(\frac{NA_\perp}{\Tilde{\omega}}).
\end{equation}
A more complete discussion is reported in Ref.~\cite{taminiau_detection_2012} where the signal under imperfect resonance condition $\tau=\tau_k(1+\Delta)$ (where $\tau_k$ is the nominal resonant value) is then $$M(T)=1-2/(1+\delta_k^2\Tilde{\omega}^2/A_\perp^2)\sin^2(N\sqrt{A_\perp^2/\Tilde{\omega}^2+\delta_k^2}/2),\qquad \delta_k=(2k+1)\pi\Delta.$$

\subsection{Effects on noise reconstruction}
\begin{figure}[htbp]
\centering \includegraphics[width=0.65\textwidth]{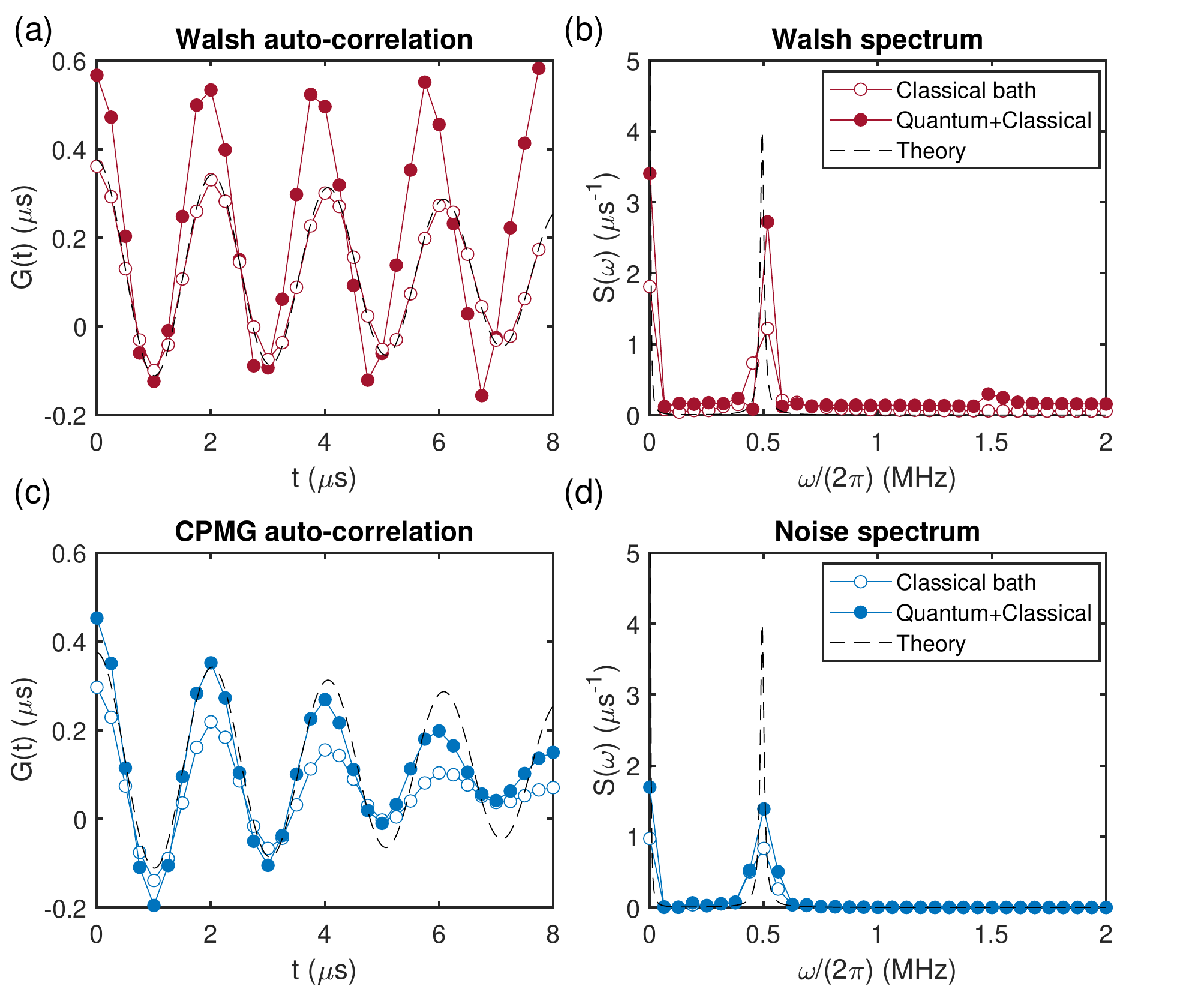}
\caption{\label{Supp_Fig_QuantumClassical} Simulation of the effects of a quantum-classical bath. The sequence parameters are $T=8\ \mu$s, $N=32$. (a)-(b) Walsh reconstruction of the auto-correlation and noise spectrum. A quantum nuclear spin with $\omega_L=-(2\pi)0.4926$~MHz, $A_\parallel=(2\pi)0.052$~MHz, $A_\perp=(2\pi)0.096$~MHz, as well as a classical noise bath with two OU components with $b^2_{0,1}=0.125,0.25\ \mu$s, $\tau_{c0,c1}=80,16\ \mu$s at frequencies $0,\omega_L$, respectively, form the quantum-classical bath. In comparison, a classical bath without the quantum nuclear spin is also simulated and plotted. The theoretical curves for the classical noise are given by $G(t)=b_0^2e^{-t/\tau_{c0}}+b_1e^{-t/\tau_{c1}}\cos(\omega_L t)$ and $S(\omega)=2{b_0^2\tau_{c0}}/(1+\omega^2\tau_c^2)+{b_1^2\tau_{c1}}/(1+(\omega-\omega_L)^2\tau_c^2)+{b_1^2\tau_{c1}}/({1+(\omega+\omega_L)^2\tau_{c1}^2})$ are shown in dashed lines. (c)-(d) CPMG reconstruction of the auto-correlation and noise spectrum. Noise models used are the same as (a)-(b). We note that both in this figure and the main text experimental figure, the absolute values of the reconstructed spectrum are plotted (when the quantum part is introduced, negative values are obtained for some points).}
\end{figure}

For typical noise reconstruction with periodic sequences and tunable time, one can always reconstruct the classical and the quantum part by sweeping the number of sequences and fitting the result to Eq.~\eqref{Supp_eq:Signal}. The quantum effects can then be subtracted from the (classical) noise spectroscopy results. Here we focus on discussing the error of classical noise reconstruction introduced by the unknown quantum part when the sequence time is fixed (or limited by hardware, etc.) and the Walsh and CPMG schemes are utilized with the assumption of a classical noise such that the signal formula ${\cal S}=(1+e^{-\Tilde\chi})/2$ is still used. 

We simulate the reconstruction of the auto-correlation and spectrum with the same experimental sequences introduced in the main text. To better mimic the situation in the experiment, we use a quantum-classical noise model comprised of a quantum nuclear spin and a classical spin bath. The classical spin bath includes a zero frequency OU noise with a correlation time $\tau_{c0}=80\ \mu$s and a shifted OU noise with the Larmor frequency of the $^{13}$C nuclear spin and a correlation time $\tau_{c1}=16\ \mu$s. In Fig.~\ref{Supp_Fig_QuantumClassical}, we compare the reconstruction with and without the existence of the quantum spin. We see that the quantum spin introduces additional errors in the auto-correlation reconstruction for both the Walsh and CPMG methods. In particular, while Walsh reconstruction for a pure classical model is almost perfect, the quantum contribution leads  to a worse reconstruction than the CPMG method, which instead is closer to the theoretical predictions. Our simulation matches the experimental observation in the main text and clearly shows the effects of the quantum noise in a quantum-classical model. 

In Fig.~\ref{Supp_Fig_CPMG}(a) we further use the same noise model to simulate a series of CPMG measurements, as a comparison to the experimental measurement shown in Fig.~\ref{Supp_Fig_CPMG}(b). Both simulation and experiment show coherent oscillations on the two sides of the Larmor frequency ($\sim0.5$~MHz), and fast decay is seen at the Larmor frequency. The discrepancy between the simulation and experiment could be due to the imperfection in the control pulse and different size and strength of the actual qubit environment.

\begin{figure}[htbp]
\centering \includegraphics[width=\textwidth]{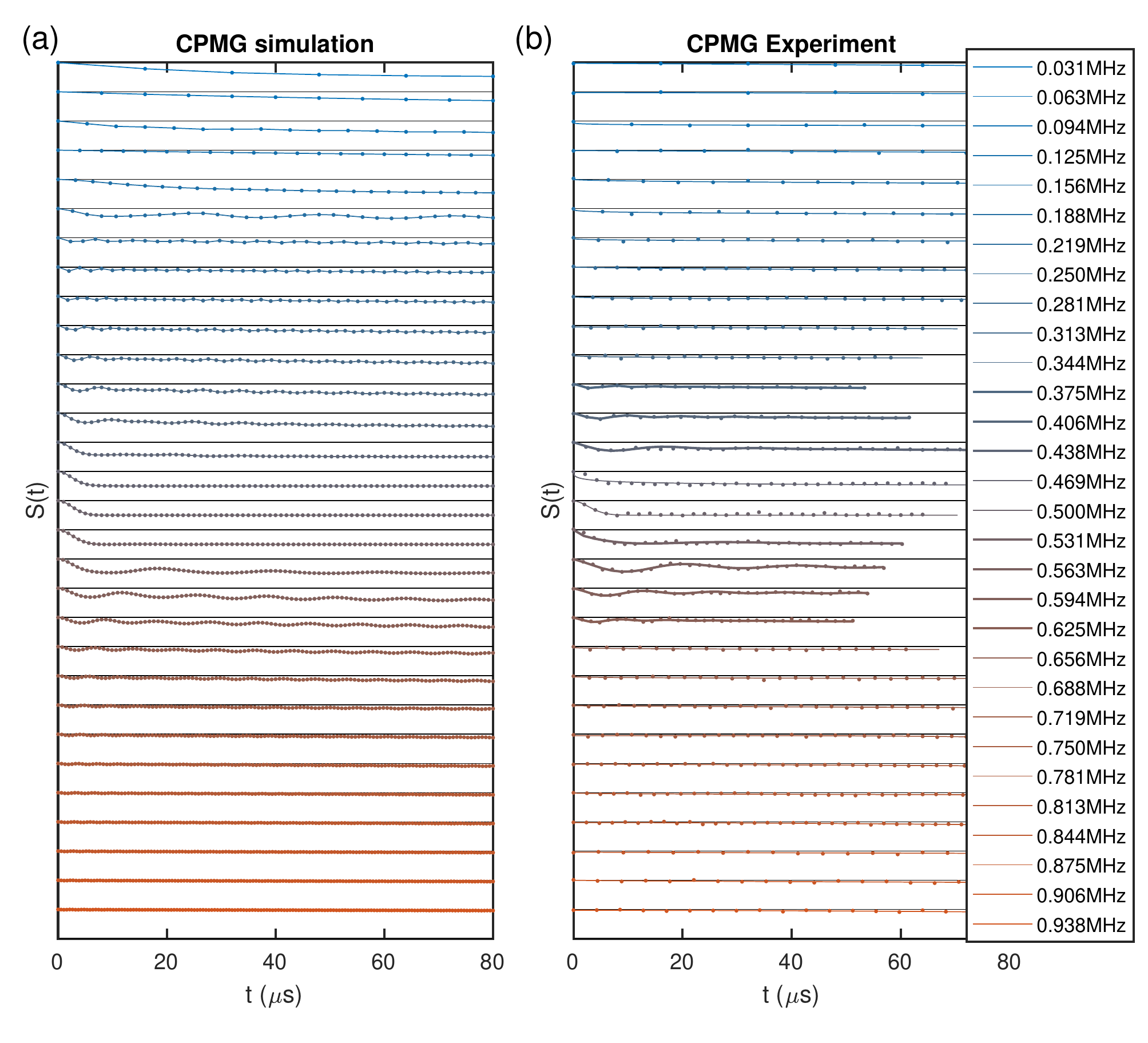}
\caption{\label{Supp_Fig_CPMG} CPMG simulation and experiment. (a) CPMG simulation under a quantum-classical noise with two quantum spins with parameters $\omega_L=-(2\pi)0.4926$~MHz, $A_{\parallel 1}= (2\pi)37\text{ kHz},\ A_{\perp 1}= (2\pi)116\text{ kHz}$ and $A_{\parallel 2}= -(2\pi)48\text{ kHz},\ A_{\perp 2}= (2\pi)92\text{ kHz}$, as well as a classical noise bath with two OU components with $b^2_{0,1}=0.125,0.25\ \mu$s, $\tau_{c0,c1}=80,16\ \mu$s at frequencies $0,\omega_L$, respectively.  (b) CPMG experiment under different inter-pulse delay $\tau$. The corresponding (first order) noise component frequencies $\pi/\tau$  are indicated in the legend.}
\end{figure}


To overcome the distortion that a quantum environment introduces in the reconstructed noise spectrum, a simple strategy is to use periodic sequences to first determine the parameters of the quantum environment. Then, the correction to the signal can be easily calculated numerically for each Walsh sequence, which allows determining the decay $\chi$ due to the classical noise only, and thus correctly reconstructing the noise auto-correlation.

\clearpage

\section{Experimental details}
\subsection{Setup and control}

We use a homemade confocal microscope to apply a $532$nm laser (SPROUT, Lighthouse Photonics), and detect the fluorescence with a single-photon counting module (Perkin Elmer SPCM-AQRH-14). A permanent magnet applies a static magnetic field at $\sim$460~G along the NV axis to lift the degeneracy of the NV ground states. Two ground states $|m_S=0\rangle$ and $|m_S=-1\rangle$ are used as the spin qubit. The magnetic field is tuned to excited state level anti-crossing (ESLAC) such that the nuclear spin of $^{14}$N is polarized to $\ket{m_I=+1}$ state under the illumination of a green laser~\cite{jacques_dynamic_2009}. A $1.5$~GHz microwave signal generated by a RF signal generator (Stanford Research System) is mixed with a phase-controlled $50$~MHz signal generated by an arbitrary waveform generator (WX1284C), which is subsequently amplified by a microwave amplifier (Mini-Circuits, 
ZHL-30W-252-S+) before applying to the diamond by a $25$~$\mu$m diameter straight copper wire. All these optical and electronic instruments are synchronized by a pulse blaster (PulseBlasterESR-PRO 500). 

The noise spectrum dominated by the nuclear spin of $^{13}$C atoms surrounding a single NV center at around $\sim30\mu m$ depth in the diamond(Fig.~\ref{Supp_Fig_Experiment}(a)) is reconstructed by both CPMG and Walsh methods.
In both Walsh and CPMG experiments, we use Gaussian-shaped pulse defined by
\begin{equation}
    V(t)=V_0\cdot e^{-\frac{1}{2}\frac{(t-t_0)^2}{(6\Delta t)^2}}\cdot \sin{(2\pi ft +\phi)},
    \label{Supp_eq:GaussianSine}
\end{equation}
where $V_0$, $f$, $\phi$ are the amplitude, frequency and phase of the pulse,  and $\Delta t$, $t_0$ are the ``duration" and time location of the pulse. 
A $9.605$~MHz Rabi frequency is used for control pulses in experiments in both the main text and the supplement. 
For both Ramsey and spin echo experiments to characterize the $T_2^*$ and $T_2$  shown in Fig.~\ref{Supp_Fig_Experiment}(b,c), we modulate the phase $\phi=df\cdot t$ [Eq.~\eqref{Supp_eq:GaussianSine}] of the last $\pi/2$ pulse  with the sequence time $t$ to obtain the oscillation signals.

\begin{figure}[htbp]
\centering \includegraphics[width=\textwidth]{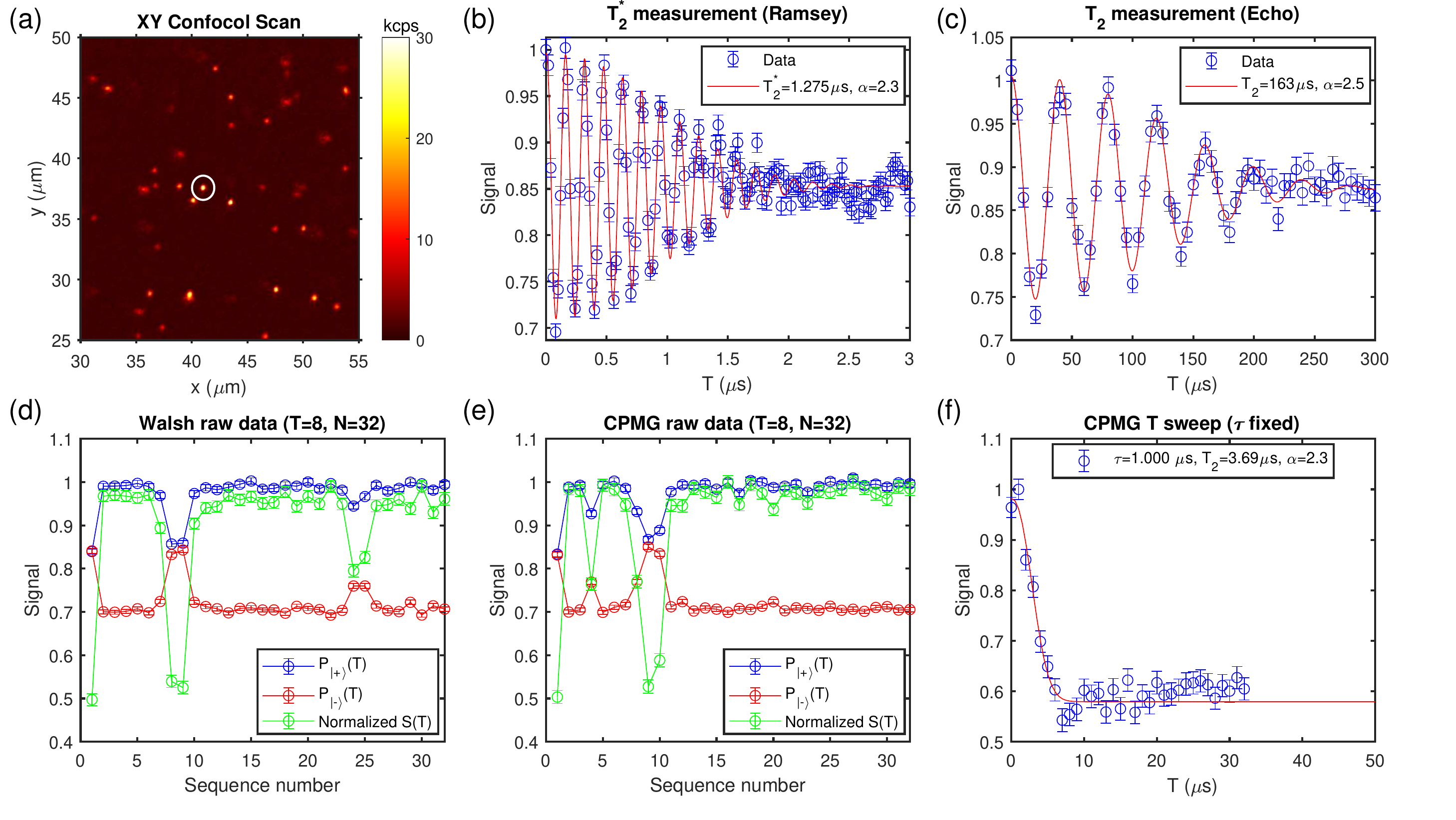}
\caption{\label{Supp_Fig_Experiment} (a) XY confocal scan. The used NV center in this work is highlighted by the white circle. (b) Ramsey measurement. We modulate the last $\pi/2$ pulse with a frequency $df=6.25$~MHz to visualize coherence oscillations. We use a formula $S(t)=c_0+c_1\cos(\omega t)e^{-(t/T_2)^\alpha}$ to fit the experimental data and obtain the coherence time $T_{2}^*=1.275\ \mu$s. Experimental average number is $10^6$.  (c) Hahn echo measurement. The last $\pi/2$ pulse is modulated with a frequency $df=0.025$~MHz and the coherence time is obtained as $T_2=147\ \mu$s by fitting to the same model in (b). Experimental average number is $10^6$.  (d) Raw data for Walsh experiment in the main text. Horizontal axis represents the sequence number. The projective measurement results $P_{\ket{+}}(t)$, $P_{\ket{-}}(t)$ and the normalized data are plotted in blue, red and green curves. The sequence time $T=8\ \mu$s and the experimental average number is $4\times10^6$. (e) Raw data for CPMG experiment in the main text. Other experimental parameters are the same as (d). (f) An exemplary CPMG experiment with fixed inter-pulse delay $\tau=1\ \mu$s such that the resonant frequency component is at $0.5$~MHz. The sequence time $T$ is swept by varing the pulse number. We note that no phase modulation is used in the experiments in (d-f).  }
\end{figure}

\subsection{Data processing} 

To suppress pulse errors and eliminate the effects due to temperature or mechanical drifts in data collection, we use the following procedure to measure and process the experimental data. In both experiments, the phase of all spin-flip $\pi$-pulses has a $90^{\circ}$ shift relative to the first $\pi/2$-pulse such that the qubit is optimally protected (similar to a spin-locking condition~\cite{wang_coherence_2020}). The last $\pi/2$ pulse has either the same phase or a $180^\circ$ phase shift with respect to the first $\pi/2$ pulse such that either $P_{\ket{-}}$ and $P_{\ket{+}}$ is obtained from the normalized fluorescence data (divided by reference signal, see main text Fig. 3(b)). For each experiment, we measure both $P_{\ket{+}}$ and $P_{\ket{-}}$ (the blue and red line in Fig.~\ref{Supp_Fig_Experiment}(d,e)) and use the equality $P_{\ket{+}}=1/2+(P_{\ket{+}}-P_{\ket{-}})/2$ to suppress the common-mode noise. Taking into account the signal contrast $c$ (0.3 for our experiment), the way to extract the normalized value for $P_{\ket{+}}$ from practical measurement outcomes $P_{\ket{\pm}}^\prime$ are through the relation
\begin{equation}
   P_{\ket{+}}=\frac12+\frac12\frac{({P_{\ket{+}}^\prime-P_{\ket{-}}^\prime })/c}{({P_{\ket{+}}^\prime+P_{\ket{-}}^\prime})/({2-c})}=\frac12+\frac{2-c}{2c}\frac{{P_{\ket{+}}^\prime-P_{\ket{-}}^\prime }}{{P_{\ket{+}}^\prime+P_{\ket{-}}^\prime}}.
   \label{contrast}
\end{equation}
The normalized data acquired from Eq.~\eqref{contrast} is shown as the green lines in Figs.~\ref{Supp_Fig_Experiment}(d,e).

\subsection{Error propagation} 
\begin{figure}[h]
\centering \includegraphics[width=0.85\textwidth]{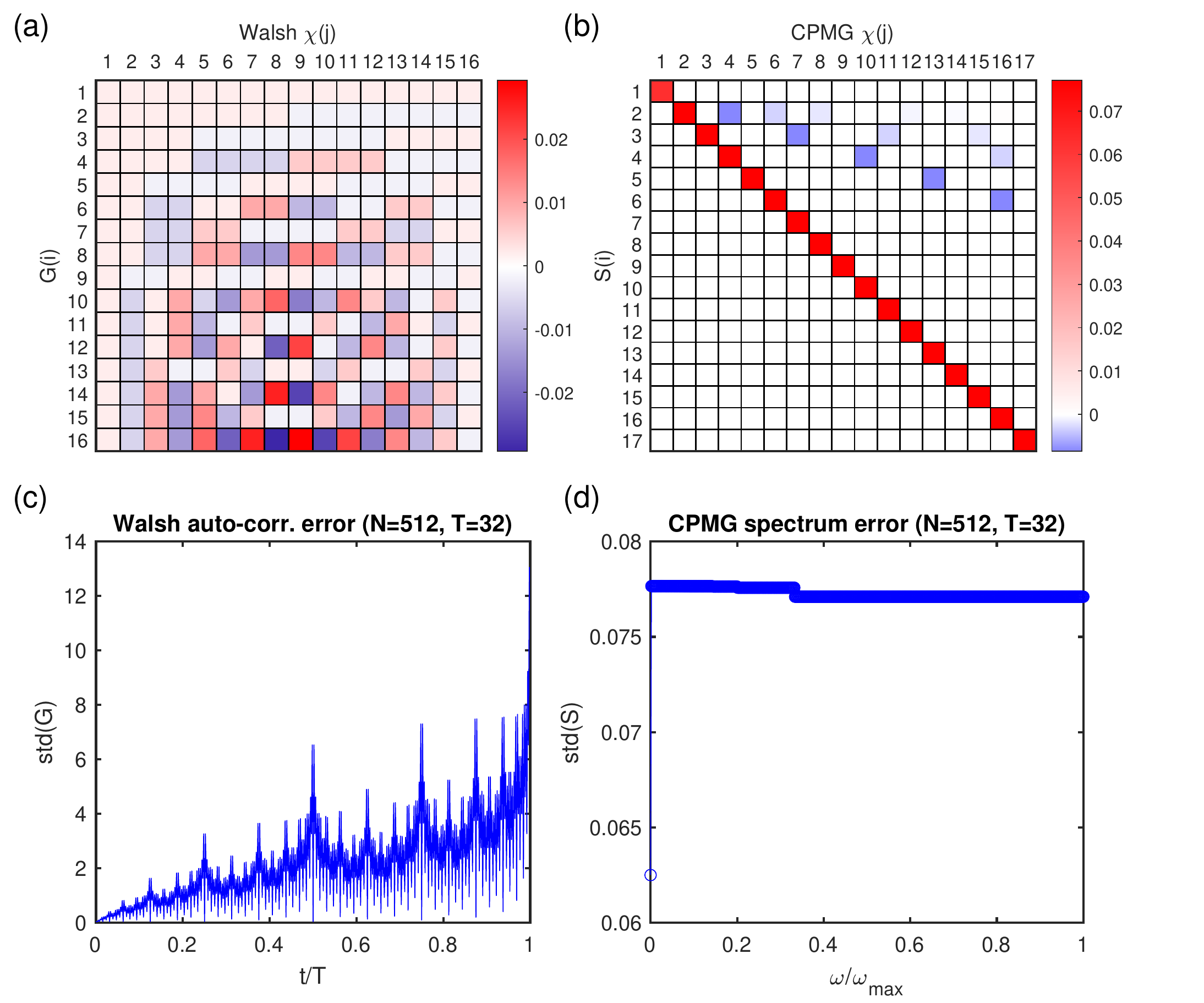}
\caption{\label{Supp_Fig_Transform} Error propagation. (a) Transform matrix from $\chi(j)$ to $G(i)$ for Walsh reconstruction method. (b) Transform matrix from $\chi(j)$ to $S(i)$ for CPMG reconstruction method. For both plots in (a) and (b), parameters $N=16,T=32$ are used. (c) Auto-correlation reconstruction error for Walsh method. We assume the same measurement uncertainty $\sigma_{\chi_k}=1$ for all $\chi_k$ and calculate the propagated error $\sigma_{G}$ using Eq.~\eqref{Supp_eq:Walsherror}. (d) Noise spectrum reconstruction error for CPMG method. We assume the same measurement uncertainty $\sigma_{\chi_k}=1$ for all $\chi_k$ and calculate the propagated error $\sigma_{S}$ using Eq.~\eqref{Supp_eq:CPMGerror}. }
\end{figure}

As shown in Figs.~\eqref{Supp_Fig_Experiment}(d,e), the measurement uncertainties for both the Walsh and CPMG methods are similar, while in the main text the obtained auto-correlation and spectrum for the Walsh method give much larger errorbar than the CPMG method. This is mainly due to the error propagation in the different matrix transformations for both method. Here we include a detailed discussion.

The error bar for $P_{\ket{\pm}}^\prime$ is given by the uncertainty in photon number counting taking into account the $\sim 10^6$ (depends on specific experiments) averages. The error $\sigma_\chi$ of the extracted attenuation function $\chi=-\ln(2P_{\ket{+}}-1)$ is then calculated with
\begin{equation}
\sigma_\chi=\frac{2\sigma_{P_{\ket{+}}}}{2 {P_{\ket{+}}}-1},
   \label{Supp_eq:error2}
\end{equation}
where the error of the normalized $P_{\ket{+}}$ is obtained through
\begin{equation}
   \sigma_{P_{\ket{+}}}=\frac{2-c}{2c}\cdot \sqrt{\frac{(2\sigma_{P_{\ket{-}}^\prime} P_{\ket{+}}^\prime)^2+(2\sigma_{P_{\ket{+}}^\prime} P_{\ket{-}})^2}{(P_{\ket{+}}^\prime+P_{\ket{-}}^\prime)^2}}.
\end{equation}
Based on the error in $\chi$, we then obtain the error of the reconstructed auto-correlation and noise spectrum for both the Walsh and the CPMG methods.
We assume the transformation from $\chi$ to $S$ or $G$ are represented by a matrix $M$, the error of each point in the $S$ or $G$ can be calculated with
\begin{equation}
   \sigma_{S_i (G_i))}=\sqrt{\sum_{k} |M_{i,k}{\sigma_\chi}_k|^2}.
   \label{Supp_eq:CPMGerror}
\end{equation}
In particular, the error propagation from experimental $\chi$ to the auto-correlation function in Walsh experiment is calculated with
\begin{align}
   \sigma_{G_i}&=\sqrt{\sum_m(\frac{\partial{\sum_kT_N^{-1}(k,i)D_N^{-1}(k,k)\frac{2N}{T^2}\sum_m W_N^{-1}(k,m)\chi_m}}{\partial \chi_m}{\sigma_\chi}_m)^2}\\
   &=\frac{2N}{T^2}\sqrt{\sum_m(\sum_kT_N^{-1}(k,i)D_N^{-1}(k,k) W_N^{-1}(k,m){\sigma_\chi}_m)^2}, 
   \label{Supp_eq:Walsherror}
\end{align}
where $T_N^{-1}$,$D_N^{-1}$, and $W_N^{-1}$ are defined in Eqs.~\eqref{walshchi2L} and \eqref{walshL2G}.

The noise spectrum for CPMG is reconstructed from $\chi_i$ through an almost diagonal matrix transform even when the higher order harmonics are taken into account; however, the auto-correlation for Walsh is reconstructed from $\chi_i$ through a complicated non-diagonal matrix as shown in Figs.~\ref{Supp_Fig_Transform}(a-b). Thus, even a few points with large error would contaminate most points in the reconstructed curve, which become a potential limitation for practical applications. In Figs.~\ref{Supp_Fig_Transform}(c-d), we evaluate such an effect by inputting the same error to measured $\sigma_{\chi_m}\equiv1$ and calculating the error in auto-correlation (spectrum) for Walsh (CPMG) reconstruction. For Walsh method, the error in $G(t)$ increases with $t$, while for the CPMG method the error in $S(\omega)$ almost does not depend on the value of frequency $\omega$.

\subsection{CPMG experiment with fixed $\tau$} 

As mentioned in the discussion in Sec.~\ref{Supp_DerivationQCmodel}, the typical way to reconstruct a quantum-classical noise is to measure the time evolution $P_{\ket{+}}(T)$ under CPMG sequences with different inter-pulse dalay $\tau$ and fitting to the analytical formula in Eq.~\eqref{Supp_eq:Signal}. Here we choose a series of the $\tau$ such that the measured noise frequencies taking values in
$\left\{\frac{0.5}{16},\frac{1}{16},\dots,\frac{14.5}{16},\frac{15}{16}\right\}\text{MHz}$ and plot
the experiment result in Fig.~\ref{Supp_Fig_CPMG}(b). When far away from the $^{13}$C Larmor frequency (-0.496~MHz), the evolution can be well fitted by an exponential decay function, 
$P_{\ket{+}}(N)=\frac{1}{2}(1+ce^{-(\frac{N\tau}{T_2})^\alpha})$, where $c$ is a phenomenological fitting parameter to better capture the overall especially the long-time decay feature.
When the resonance frequency is around the $^{13}$C Larmor frequency, coherent oscillations are observed, the data are fitted with a modified quantum-classical model formula 
\begin{equation}
   P_{\ket{+}}(N)=\frac{1}{2}\left[1+e^{-(\frac{N\tau}{T_2})^\alpha}(1-2e^{-(\frac{N\tau}{T_{2\rho}})^{\beta}}\frac{A_\perp^2}{\Tilde{\omega}^2}\sin^2\frac{\phi_0}{2}\sin^2\frac{\phi_1}{2}\frac{\sin^2\frac{N\varphi}{2}}{\cos^2\frac{\varphi}{2}})\right],
   \label{CPMGExp_CQ}
\end{equation}
where $e^{-(\frac{N\tau}{T_{2\rho}})^{\beta}}$ is a phenomenological decay factor of the oscillation part in Eq.~\eqref{Supp_eq:M(T)}, which might be due to the imperfection of the control pulses and hardware resolution (the AWG time steps are set to 2~ns). We note that a better pulse spacing sampling for better frequency resolution can be achieved with quantum interpolation~\cite{ajoy_quantum_2017,liu_quantum_2019}. A series of different (but close) hyperfine interaction parameters $A_{\parallel}$ and $A_{\perp}$ are obtained when the fitting is applied for data corresponding to frequencies $\frac{1}{2\tau}=\left\{\frac{6}{16},\frac{6.5}{16},\ldots,\frac{10}{16}\right\}$~MHz. We classify the fitting results to two sets of hyperfine strengths on the two sides of the $^{13}$C Larmor frequency $A_{\parallel 1}= (2\pi)37\pm16\text{ kHz},\ A_{\perp 1}= (2\pi)116\pm20\text{ kHz}$ and $A_{\parallel 2}= -(2\pi)48\pm14\text{ kHz},\ A_{\perp 2}= (2\pi)92\pm19\text{ kHz}$, which indicates that there are possibly two distinct quantum spins in the bath. We note that here the experimental results might be due to more quantum spins with small coupling strengths. However, they are hard to distinguish due to fast signal decay, and fully characterizing the number of quantum spins and their hyperfine coupling strengths are not the focus of this work. As shown in Fig.~\ref{Supp_Fig_CPMG}, the simulation with two quantum spins successfully reproduce the experimental results qualitatively.

\end{widetext}

\end{document}